\documentclass[letterpaper,11pt]{article}

\usepackage{titlesec}
\usepackage[utf8]{inputenc}
\usepackage{graphicx,fullpage,paralist}
\usepackage{setspace}
\usepackage{caption}
\usepackage{amsmath, amssymb, amsthm}
\usepackage{comment,hyperref}
\usepackage{thmtools}
\usepackage{tikz}

\usepackage{pgfplots}
\usepackage{algorithm}
\usepackage[algo2e]{algorithm2e}
\usepackage{lscape}
\usepackage{float}
\usepackage[toc,page,header]{appendix}
\usepackage{minitoc}
\usepackage{varwidth}
\pgfplotsset{compat=1.18,
 every axis/.append style={
 axis x line*=bottom,
 axis y line*=left,
 axis line style={-},
 tick style={black}
 },
 every axis plot/.append style={-}
}
\usepgfplotslibrary{fillbetween}
\usetikzlibrary{backgrounds}
\usetikzlibrary{patterns}
\usepackage{natbib}
\usepackage{mathtools}

\usepackage{amsbsy}

\newtheorem*{ra3'}{RA 3'}
\newtheorem*{ra4'}{RA 4'}
\newtheorem*{ra3''}{RA 3''}
\newtheorem*{ra4''}{RA 4''}

\usepackage{lipsum}
\usepackage{multirow}
\usepackage[noend]{algpseudocode}

\newlength{\algofontsize}
\usepackage[textsize=tiny,textwidth=2cm]{todonotes}

\usepackage{xcolor}
\hypersetup{
	colorlinks,
	linkcolor={red!50!black},
	citecolor={blue!50!black},
	urlcolor={blue!80!black}
}

\renewcommand{\mathbf}{\boldsymbol}

\usetikzlibrary{shapes,decorations,arrows,calc,arrows.meta,fit,positioning,automata}
\tikzset{
 auto,node distance =1 cm and 1 cm,semithick,
 state/.style ={ellipse, draw, minimum width = 0.7 cm,rounded corners},
 point/.style = {circle, draw, inner sep=0.04cm,fill,node contents={}},
 bidirected/.style={Latex-Latex,dashed},
 el/.style = {inner sep=2pt, align=left, sloped}
}
\usepackage{makecell}

\usepackage{subcaption}
\usepackage{graphicx}
\usepackage{booktabs}
\newcommand{\addauthor}[2][]{}

\usepackage{textcomp}
\usepackage{breqn}
\graphicspath{{../}{./}}

\titleformat{\paragraph}[runin]{\normalfont\normalsize\bfseries}{\theparagraph}{0.5em}{}

\begin{document}

\doparttoc
\faketableofcontents

\algrenewcommand\algorithmicrequire{\textbf{Input:}}
\algrenewcommand\algorithmicensure{\textbf{Output:}}

\title{
 Residential Battery Pooling Under Backup Commitments
}
\author{
Jerry Anunrojwong\\
\small Clean Energy Institute, University of Washington\\
\small \texttt{janunroj@uw.edu}
\and
Baosen Zhang\\
\small Department of Electrical and Computer Engineering, University of Washington\\
\small \texttt{zhangbao@uw.edu}
}
\date{}

\maketitle

\begin{abstract}
Residential batteries increasingly serve two roles: they can earn money by arbitraging wholesale prices and providing grid services, and they provide backup power during outages. This dual use creates a basic tradeoff between earning market value and preserving outage readiness. Coordination across many batteries can help, but a provider cannot treat the fleet as a single virtual battery when each household is promised its own backup protection.

We compare standalone control, in which each home is dispatched independently, with pooling, in which homes are coordinated while each battery retains its own state of charge and household-specific backup requirement. Both regimes are implemented as model predictive control problems with 15-minute decision intervals and evaluated using household telemetry together with ERCOT market inputs. The empirical design focuses on the 543 homes in our sample that can support at least one backup product in standalone operation and studies backup caps ranging from 2 to 24 hours. Lower caps relax backup obligations, while the 24-hour cap coincides with assigning each home its own longest feasible backup tier.

Pooling remains beneficial in this service-constrained setting, but its value declines smoothly as backup obligations tighten. Standalone firm margin ranges from \$11.06 per home per week at the 2-hour cap to \$10.79 at the 24-hour cap, while pooling benefit falls from \$1.49 to \$1.27 per home per week. Relative to standalone firm margin, pooling is worth about 13.5\% at the 2-hour cap and about 11.8\% at the 24-hour cap. Coordination therefore still helps after preserving household-level backup guarantees, but its value declines as backup obligations tighten.
\end{abstract}

\section{Introduction}\label{sec:introduction}

Residential batteries are increasingly sold as a combined operating and resilience product. During normal conditions, a provider wants to use the battery to lower procurement costs, absorb rooftop solar, and respond to time-varying prices. During outages, the same battery is expected to keep the household served. That dual role makes residential battery management an operational problem rather than a simple arbitrage exercise: energy used for near-term economics is energy that may not be available when backup service is needed.

The same tradeoff becomes more complex when a provider manages many homes. Heterogeneity in load, solar production, battery size, state of charge, and price exposure creates room for coordination to improve performance. That coordination is economically meaningful only if the model respects the household-level backup commitments the provider is making. A provider cannot collapse a fleet of behind-the-meter batteries into one frictionless virtual device when each household expects its own battery to satisfy an explicit or implicit backup commitment. Once those household-level obligations are kept explicit, the central question is not simply whether pooling helps by smoothing differences across homes; it is how much operational value coordination can deliver under household-level service constraints.

We study that question in a setting motivated by behind-the-meter battery providers such as Base Power (\url{https://www.basepowercompany.com/}). We compare two operating regimes. In \emph{standalone} control, each home's battery is dispatched independently. In \emph{pooling}, batteries are coordinated across homes, but each home retains its own battery state and its own backup requirement. This comparison isolates the operational value of coordination relative to what the same households can achieve under standalone service.

We develop a framework for comparing standalone and pooled battery operation that preserves household-level backup commitments and quantifies the value of coordination under those constraints. We formulate standalone service as the reduced linear program solved within a model predictive control framework, with battery dynamics, reserve floors, and the routing variables needed to represent the provider’s cash flows under the tariff. We then extend that framework to pooled operation without replacing the fleet with a single aggregate battery.

Our empirical design studies a common spectrum of backup caps. We first identify the homes in our sample that can support at least one positive backup requirement in standalone operation. We then compare standalone and pooled operation across backup caps of 2, 4, 6, 8, 12, and 24 hours, assigning each home the shorter of the common cap and the longest backup duration it can support in standalone operation. Pooling benefit falls smoothly from \$1.49 per home per week at the 2-hour cap to \$1.27 at the 24-hour cap, while standalone firm margin remains close to \$11 per home per week. Relative to standalone firm margin, pooling is worth about 13.5\% at the 2-hour cap and 11.8\% at the 24-hour cap.

Taken together, these results show that residential battery pooling creates positive coordination gains even when backup requirements are enforced at the household level. Those gains decline as backup obligations tighten and remain positive across the full range of backup caps we consider.

The rest of the paper proceeds as follows. Section~\ref{sec:literature} reviews the related literature. Section~\ref{sec:data} describes the empirical setting and data. Section~\ref{sec:model} presents the control model. Section~\ref{sec:backup-accounting} defines the reserve requirements. Section~\ref{sec:results} reports the main numerical results. Section~\ref{sec:conclusion} concludes.

\section{Related Literature}\label{sec:literature}

This paper sits at the intersection of storage operations, behind-the-meter battery control, and backup-oriented battery coordination. Its central positioning point is that it studies coordination across residential batteries when backup commitments remain attached to individual homes. In that setting, pooling is not well represented by a virtual-battery abstraction: portfolio-level dispatch can coordinate energy use across homes, but battery states and reserve requirements remain household specific.

A first related literature studies how storage should be operated, bid, or coordinated when decisions are intertemporal and uncertainty matters. In operations management and operations research, this includes work on storage commitments and control under uncertainty \citep{kim2011optimal, cruise2019control, lohndorf2023coordination}. This work provides the dynamic-control backdrop for our problem: stored energy has value because it can be shifted across time, but that value depends on uncertain prices, loads, and future constraints.

Closely related papers study the economics of storage placement, aggregation, strategic operation, adoption, and service stacking. This includes work on the investment and operating tradeoffs between distributed and centralized storage \citep{wu2023distributedstorage}, the efficiency and market-power implications of DER aggregation and battery operation \citep{gao2024aggregating, anunrojwong2025battery}, the economics of residential battery adoption \citep{kaps2025residential}, and the profitability of stacking multiple services on a battery \citep{bae2025energystacking}. These papers provide the economic backdrop for our question, but they typically do not ask how much coordination is worth once household-level outage-protection promises must also be preserved. Our paper studies that margin directly by evaluating pooling under explicit household-specific reserve floors and measuring value in a firm-margin framework tied to the backup product being sold.

A second literature in engineering develops optimization and MPC methods for behind-the-meter PV-battery systems and home energy management. These papers study residential or building-scale control under tariffs, forecast uncertainty, self-consumption objectives, and battery degradation \citep{sun2016nonlinear, cai2019aging, wu2022hems, zou2023pvbmpc}. Related work compares centralized and distributed storage operation or examines how multiple applications can be combined within a residential battery \citep{zakeri2021centralized, parra2019combining}. Some related work also studies backup readiness, outage resilience, and the tradeoff between bill savings and backup power in residential or distributed storage systems \citep{khodaei2014resiliency, tobajas2022resilience, gorman2025backup}. This work is relevant because backup service changes the operational meaning of stored energy: some energy is protected inventory rather than discretionary arbitrage capacity. Our standalone controller is closest in spirit to this literature, but the comparison we study is different. Rather than comparing alternative controllers for a single home, we compare standalone operation with pooling across many homes while keeping each home's battery state and household-specific backup requirement explicit.

\section{Empirical Setting and Data}\label{sec:data}

The empirical analysis focuses on a single operating week, August 1--7, 2025. We evaluate standalone control and pooling over that week using observed household and market data. The single-week design is intended as a controlled comparison across operating regimes rather than a seasonal estimate of annual profitability; we therefore focus on differences between controllers evaluated on the same homes, prices, and reserve rules. Base Power provides the household telemetry and battery metadata used in the numerical experiments, including 1-minute measurements of household load and solar production together with home-level battery characteristics such as usable energy capacity and charge and discharge limits. The full sample contains 581 homes with complete data over the week. The main sample in the paper is the 543-home retained cohort obtained by rebuilding feasibility under our reserve construction and dropping homes that are infeasible even at the shortest 2-hour backup requirement. ERCOT provides the wholesale price series used in the analysis. All control problems are solved at 15-minute intervals.

\begin{figure}[t]
\centering
\includegraphics[width=0.62\textwidth]{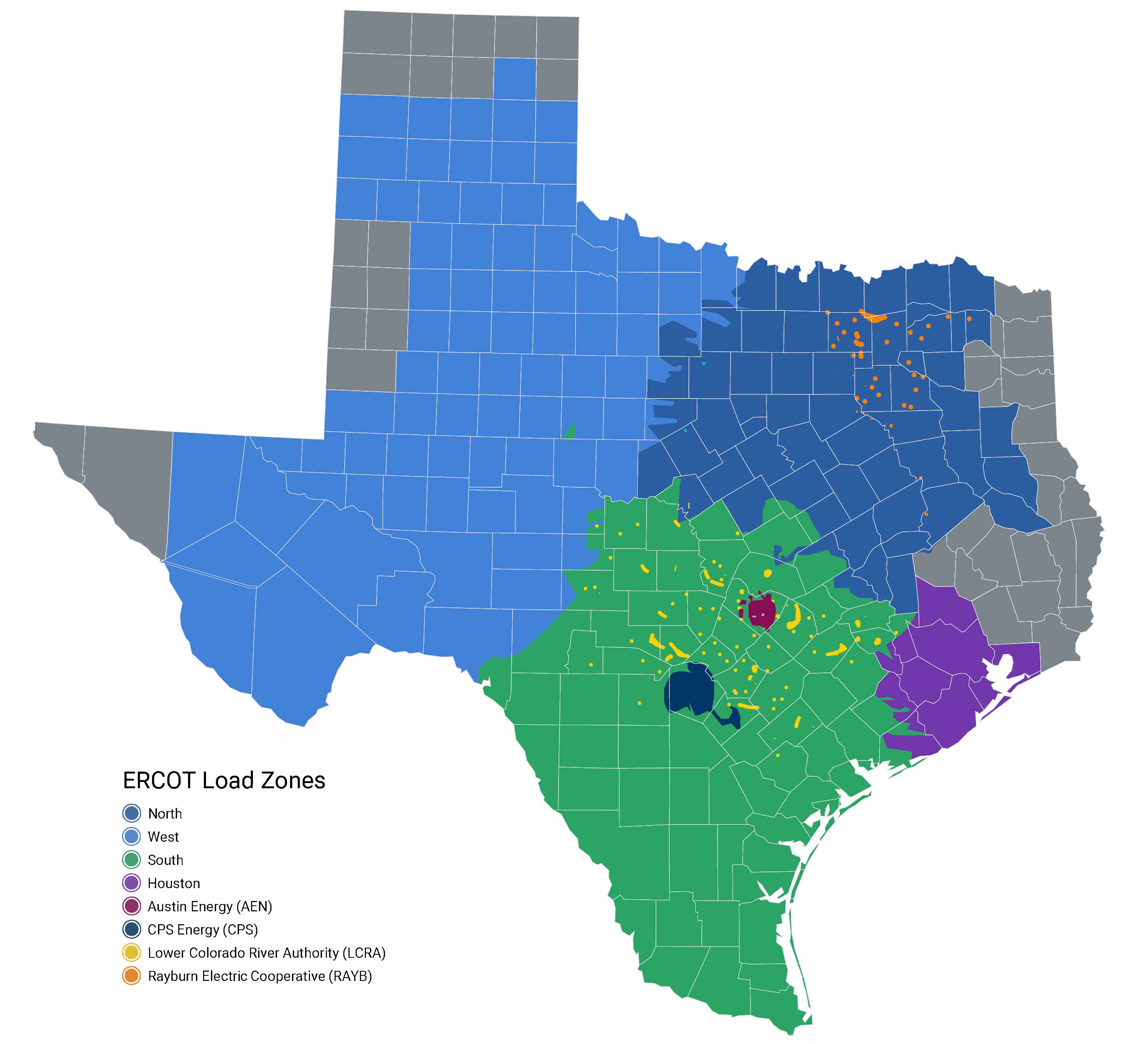}
\caption{ERCOT load zones. The empirical analysis uses ERCOT real-time prices at settlement point \texttt{LZ\_SOUTH}, the load zone containing the Base homes in our sample.}
\label{fig:ercot-zones}
\end{figure}

\begin{figure}[t]
\centering
\begin{minipage}[t]{0.48\textwidth}
\centering
\begin{tikzpicture}
\begin{axis}[
width=\textwidth,
height=0.24\textheight,
title={Home A: typical non-solar},
xmin=1, xmax=672,
ymin=-6, ymax=18,
ytick={0,5,10,15},
xtick={1,97,193,289,385,481,577},
xticklabels={Aug 1,Aug 2,Aug 3,Aug 4,Aug 5,Aug 6,Aug 7},
xticklabel style={font=\scriptsize,rotate=45,anchor=east},
ylabel={Net load (kW)},
grid=both,
major grid style={gray!20},
minor grid style={gray!10}
]
\addplot+[mark=none, line width=0.7pt] coordinates {(1,7.344) (2,6.643) (3,5.660) (4,4.539) (5,3.481) (6,3.289) (7,3.141) (8,2.868) (9,2.876) (10,2.839) (11,2.932) (12,2.937) (13,2.758) (14,2.834) (15,2.798) (16,3.426) (17,2.913) (18,2.879) (19,2.302) (20,2.829) (21,3.330) (22,3.353) (23,2.021) (24,2.479) (25,1.388) (26,2.651) (27,1.455) (28,1.859) (29,2.121) (30,0.963) (31,2.460) (32,1.085) (33,2.311) (34,1.057) (35,1.862) (36,1.207) (37,1.607) (38,1.374) (39,1.409) (40,1.402) (41,1.210) (42,1.506) (43,0.863) (44,1.763) (45,0.541) (46,0.498) (47,0.517) (48,0.454) (49,0.531) (50,0.496) (51,0.569) (52,0.641) (53,0.674) (54,0.705) (55,0.688) (56,0.737) (57,0.663) (58,0.709) (59,0.638) (60,1.926) (61,2.268) (62,2.807) (63,1.807) (64,1.899) (65,2.044) (66,0.766) (67,0.657) (68,1.041) (69,3.022) (70,2.074) (71,0.568) (72,2.617) (73,2.987) (74,1.381) (75,0.958) (76,3.044) (77,3.047) (78,2.826) (79,0.529) (80,2.498) (81,3.108) (82,1.469) (83,2.402) (84,2.956) (85,3.057) (86,1.174) (87,3.126) (88,3.100) (89,2.917) (90,2.953) (91,2.879) (92,2.909) (93,2.831) (94,2.866) (95,2.933) (96,2.958) (97,3.057) (98,3.135) (99,3.234) (100,2.963) (101,2.884) (102,2.953) (103,3.942) (104,5.070) (105,4.888) (106,3.549) (107,2.717) (108,2.567) (109,2.693) (110,1.294) (111,0.429) (112,1.593) (113,2.670) (114,2.558) (115,0.467) (116,0.871) (117,2.597) (118,1.111) (119,0.484) (120,1.579) (121,2.580) (122,0.650) (123,0.464) (124,0.722) (125,2.508) (126,1.182) (127,2.174) (128,2.605) (129,0.549) (130,1.596) (131,1.554) (132,0.632) (133,2.539) (134,0.479) (135,1.501) (136,1.333) (137,0.673) (138,2.179) (139,0.445) (140,0.450) (141,0.448) (142,0.455) (143,0.368) (144,0.471) (145,0.399) (146,0.457) (147,0.361) (148,0.422) (149,0.529) (150,0.516) (151,0.505) (152,1.066) (153,0.945) (154,2.956) (155,6.383) (156,4.898) (157,4.165) (158,5.924) (159,5.321) (160,5.321) (161,2.878) (162,3.639) (163,3.358) (164,3.040) (165,3.000) (166,2.689) (167,2.171) (168,3.448) (169,3.205) (170,3.208) (171,3.141) (172,1.445) (173,3.164) (174,3.063) (175,2.962) (176,1.102) (177,2.013) (178,2.952) (179,1.689) (180,2.085) (181,2.375) (182,1.416) (183,2.332) (184,1.207) (185,2.868) (186,0.751) (187,2.460) (188,1.595) (189,2.865) (190,3.208) (191,3.014) (192,3.064) (193,3.899) (194,3.059) (195,3.050) (196,2.927) (197,2.320) (198,3.000) (199,1.599) (200,2.404) (201,1.916) (202,1.827) (203,2.732) (204,0.498) (205,2.429) (206,1.516) (207,1.208) (208,2.474) (209,0.639) (210,2.419) (211,2.298) (212,1.586) (213,2.524) (214,0.792) (215,3.154) (216,1.514) (217,2.033) (218,1.742) (219,1.561) (220,1.909) (221,0.567) (222,2.345) (223,0.564) (224,1.993) (225,0.958) (226,1.332) (227,1.548) (228,0.721) (229,0.573) (230,0.596) (231,0.540) (232,0.550) (233,0.542) (234,0.564) (235,1.405) (236,2.263) (237,0.519) (238,0.539) (239,0.592) (240,0.525) (241,0.501) (242,0.537) (243,0.600) (244,0.588) (245,0.614) (246,0.758) (247,0.755) (248,0.668) (249,0.484) (250,0.576) (251,0.635) (252,0.999) (253,0.698) (254,2.722) (255,0.562) (256,1.061) (257,1.996) (258,2.855) (259,2.958) (260,2.471) (261,1.703) (262,3.098) (263,3.166) (264,3.079) (265,1.354) (266,3.076) (267,3.103) (268,2.034) (269,1.835) (270,3.076) (271,2.922) (272,0.664) (273,1.919) (274,3.056) (275,4.039) (276,6.903) (277,5.321) (278,4.109) (279,2.804) (280,1.844) (281,3.059) (282,3.129) (283,3.100) (284,3.133) (285,3.131) (286,5.259) (287,7.798) (288,6.707) (289,4.832) (290,4.708) (291,3.305) (292,3.226) (293,3.107) (294,3.063) (295,2.951) (296,2.680) (297,2.822) (298,2.814) (299,2.712) (300,3.011) (301,2.829) (302,2.796) (303,2.708) (304,2.655) (305,2.181) (306,1.925) (307,2.766) (308,3.245) (309,2.257) (310,2.724) (311,2.973) (312,1.488) (313,2.527) (314,2.132) (315,0.967) (316,2.685) (317,0.964) (318,2.653) (319,1.275) (320,1.622) (321,2.017) (322,0.908) (323,2.635) (324,0.782) (325,1.794) (326,1.663) (327,0.801) (328,2.531) (329,0.579) (330,1.490) (331,1.405) (332,0.508) (333,0.577) (334,0.541) (335,0.496) (336,0.553) (337,0.534) (338,0.908) (339,0.797) (340,0.874) (341,0.644) (342,0.724) (343,0.697) (344,1.202) (345,1.618) (346,2.242) (347,0.836) (348,4.475) (349,3.191) (350,2.713) (351,2.307) (352,0.942) (353,0.848) (354,0.948) (355,0.915) (356,0.923) (357,0.747) (358,2.213) (359,1.781) (360,0.682) (361,0.676) (362,0.721) (363,0.672) (364,2.549) (365,1.471) (366,0.601) (367,1.692) (368,1.927) (369,0.689) (370,2.885) (371,1.004) (372,1.985) (373,1.502) (374,1.874) (375,2.105) (376,0.774) (377,2.967) (378,3.287) (379,3.170) (380,2.987) (381,3.964) (382,2.966) (383,3.033) (384,3.158) (385,2.908) (386,3.277) (387,3.047) (388,2.891) (389,3.018) (390,2.954) (391,2.872) (392,2.756) (393,1.912) (394,0.637) (395,2.716) (396,2.713) (397,1.711) (398,3.660) (399,1.267) (400,2.371) (401,2.093) (402,1.777) (403,3.366) (404,1.119) (405,2.950) (406,1.381) (407,2.020) (408,1.338) (409,1.451) (410,1.701) (411,0.733) (412,2.370) (413,0.564) (414,2.054) (415,1.055) (416,1.277) (417,1.848) (418,0.614) (419,2.229) (420,0.678) (421,1.382) (422,1.297) (423,0.693) (424,2.683) (425,2.805) (426,2.188) (427,0.881) (428,0.598) (429,0.500) (430,0.581) (431,0.524) (432,0.518) (433,0.513) (434,0.552) (435,0.454) (436,0.459) (437,0.468) (438,0.460) (439,0.389) (440,0.841) (441,0.571) (442,0.570) (443,0.955) (444,0.773) (445,0.581) (446,0.498) (447,2.342) (448,0.499) (449,0.896) (450,0.557) (451,2.277) (452,2.899) (453,0.850) (454,0.854) (455,2.969) (456,2.471) (457,0.796) (458,1.826) (459,2.867) (460,3.037) (461,0.757) (462,0.916) (463,3.018) (464,2.968) (465,3.139) (466,1.674) (467,2.328) (468,3.148) (469,2.104) (470,2.342) (471,3.045) (472,3.151) (473,3.278) (474,3.295) (475,3.284) (476,3.164) (477,3.412) (478,3.301) (479,3.199) (480,3.149) (481,3.148) (482,3.119) (483,3.104) (484,2.997) (485,3.304) (486,3.052) (487,3.032) (488,3.005) (489,2.939) (490,2.898) (491,2.968) (492,2.945) (493,2.940) (494,2.798) (495,2.854) (496,2.820) (497,3.180) (498,2.214) (499,2.112) (500,2.783) (501,0.794) (502,3.667) (503,2.110) (504,2.090) (505,2.460) (506,1.183) (507,2.671) (508,0.866) (509,2.148) (510,1.334) (511,1.836) (512,1.529) (513,1.209) (514,1.984) (515,0.539) (516,2.440) (517,0.642) (518,1.539) (519,1.575) (520,0.909) (521,2.066) (522,0.530) (523,2.245) (524,0.676) (525,0.619) (526,0.505) (527,0.621) (528,0.512) (529,0.543) (530,0.458) (531,0.467) (532,0.476) (533,0.435) (534,0.509) (535,0.465) (536,0.706) (537,1.143) (538,1.086) (539,1.177) (540,0.871) (541,1.767) (542,2.297) (543,0.879) (544,1.606) (545,1.500) (546,2.185) (547,3.170) (548,1.332) (549,3.342) (550,3.140) (551,3.149) (552,3.213) (553,3.185) (554,3.085) (555,3.154) (556,1.177) (557,3.217) (558,3.285) (559,3.033) (560,0.896) (561,1.474) (562,3.243) (563,3.363) (564,1.282) (565,3.108) (566,3.204) (567,3.119) (568,2.459) (569,3.162) (570,3.512) (571,3.180) (572,3.237) (573,3.314) (574,3.158) (575,3.191) (576,3.076) (577,3.182) (578,3.225) (579,3.166) (580,3.212) (581,3.156) (582,3.094) (583,3.149) (584,2.953) (585,2.897) (586,2.871) (587,2.965) (588,2.873) (589,2.899) (590,2.919) (591,2.857) (592,1.479) (593,4.667) (594,8.157) (595,7.629) (596,5.546) (597,3.414) (598,1.485) (599,2.713) (600,2.882) (601,2.635) (602,0.766) (603,0.713) (604,2.667) (605,2.186) (606,0.608) (607,0.703) (608,2.850) (609,2.426) (610,0.592) (611,0.630) (612,1.069) (613,2.732) (614,1.866) (615,0.576) (616,0.543) (617,0.597) (618,2.649) (619,1.580) (620,0.577) (621,0.544) (622,0.585) (623,0.474) (624,0.609) (625,0.529) (626,0.524) (627,0.557) (628,0.549) (629,0.689) (630,0.662) (631,0.685) (632,0.657) (633,2.675) (634,1.566) (635,1.196) (636,1.422) (637,2.193) (638,0.836) (639,2.529) (640,2.728) (641,0.708) (642,1.535) (643,3.443) (644,1.997) (645,1.189) (646,3.248) (647,2.998) (648,1.673) (649,0.547) (650,2.348) (651,2.990) (652,1.528) (653,0.743) (654,3.008) (655,3.049) (656,3.005) (657,3.212) (658,1.283) (659,2.870) (660,3.179) (661,2.554) (662,1.686) (663,3.068) (664,3.053) (665,3.127) (666,3.130) (667,3.106) (668,3.133) (669,3.124) (670,3.159) (671,3.097) (672,3.217)};
\end{axis}
\end{tikzpicture}
\end{minipage}
\hfill
\begin{minipage}[t]{0.48\textwidth}
\centering
\begin{tikzpicture}
\begin{axis}[
width=\textwidth,
height=0.24\textheight,
title={Home B: solar-heavy},
xmin=1, xmax=672,
ymin=-6, ymax=18,
ytick={-5,0,5,10,15},
xtick={1,97,193,289,385,481,577},
xticklabels={Aug 1,Aug 2,Aug 3,Aug 4,Aug 5,Aug 6,Aug 7},
xticklabel style={font=\scriptsize,rotate=45,anchor=east},
grid=both,
major grid style={gray!20},
minor grid style={gray!10}
]
\addplot+[mark=none, line width=0.7pt] coordinates {(1,1.841) (2,1.270) (3,2.075) (4,1.769) (5,1.480) (6,1.166) (7,2.029) (8,1.675) (9,1.151) (10,1.280) (11,1.760) (12,1.318) (13,2.089) (14,2.144) (15,2.104) (16,2.092) (17,2.138) (18,2.147) (19,2.076) (20,2.032) (21,2.117) (22,2.100) (23,2.013) (24,1.980) (25,2.085) (26,2.046) (27,1.415) (28,1.925) (29,1.764) (30,1.525) (31,0.941) (32,1.227) (33,1.591) (34,0.836) (35,0.626) (36,1.342) (37,1.132) (38,0.358) (39,1.560) (40,2.169) (41,0.510) (42,0.982) (43,0.246) (44,0.283) (45,0.219) (46,0.175) (47,0.386) (48,0.312) (49,0.263) (50,1.702) (51,0.501) (52,-0.607) (53,-1.215) (54,-1.866) (55,-2.256) (56,-2.502) (57,-2.863) (58,-3.030) (59,-2.856) (60,-2.737) (61,-3.087) (62,-4.190) (63,-3.338) (64,-4.449) (65,-4.295) (66,-4.435) (67,-3.922) (68,-4.494) (69,-4.514) (70,-3.869) (71,-3.875) (72,-3.100) (73,-4.210) (74,-3.061) (75,-3.998) (76,-3.489) (77,-3.454) (78,-3.802) (79,-2.690) (80,-3.507) (81,-1.102) (82,-2.930) (83,-2.061) (84,-1.776) (85,-2.333) (86,-1.172) (87,-0.901) (88,-0.720) (89,0.232) (90,0.544) (91,0.805) (92,0.771) (93,1.290) (94,1.009) (95,1.219) (96,1.460) (97,1.234) (98,1.879) (99,1.241) (100,2.107) (101,1.303) (102,2.158) (103,2.199) (104,3.561) (105,3.059) (106,2.655) (107,3.355) (108,2.176) (109,2.643) (110,2.785) (111,2.213) (112,2.386) (113,2.260) (114,2.718) (115,2.424) (116,2.080) (117,1.357) (118,1.951) (119,1.095) (120,1.143) (121,1.494) (122,0.956) (123,0.557) (124,1.223) (125,1.083) (126,0.284) (127,1.087) (128,1.087) (129,0.239) (130,1.128) (131,1.136) (132,0.402) (133,1.075) (134,1.271) (135,0.674) (136,0.774) (137,1.434) (138,1.193) (139,0.629) (140,1.069) (141,1.288) (142,1.032) (143,0.593) (144,1.153) (145,0.856) (146,0.113) (147,0.071) (148,-0.019) (149,0.877) (150,0.754) (151,-1.456) (152,-0.477) (153,-2.772) (154,-2.105) (155,-3.185) (156,-0.321) (157,1.126) (158,-1.575) (159,-2.053) (160,-2.400) (161,-0.821) (162,1.377) (163,-2.978) (164,-2.722) (165,-3.584) (166,-1.526) (167,-3.159) (168,-3.192) (169,-0.958) (170,-1.126) (171,0.431) (172,-0.181) (173,-2.225) (174,-3.054) (175,-2.608) (176,-3.552) (177,-2.482) (178,-0.218) (179,-3.096) (180,-3.932) (181,-1.690) (182,-1.032) (183,-0.372) (184,0.359) (185,1.062) (186,0.369) (187,0.249) (188,0.745) (189,0.920) (190,0.394) (191,0.361) (192,0.859) (193,1.153) (194,1.369) (195,1.806) (196,2.002) (197,2.058) (198,2.026) (199,2.126) (200,2.086) (201,1.992) (202,2.030) (203,2.234) (204,3.178) (205,1.991) (206,2.048) (207,1.708) (208,1.520) (209,1.856) (210,1.334) (211,0.856) (212,1.252) (213,1.401) (214,0.751) (215,0.623) (216,1.217) (217,2.084) (218,0.550) (219,0.826) (220,1.545) (221,0.684) (222,1.954) (223,0.953) (224,0.529) (225,0.277) (226,0.202) (227,0.167) (228,0.097) (229,0.223) (230,0.233) (231,0.173) (232,0.110) (233,0.196) (234,0.196) (235,0.098) (236,0.126) (237,0.202) (238,0.156) (239,0.206) (240,0.169) (241,0.117) (242,-0.108) (243,-0.361) (244,-0.624) (245,-1.039) (246,-1.515) (247,-1.970) (248,-2.322) (249,-2.760) (250,-3.170) (251,-3.413) (252,-3.720) (253,-3.724) (254,-3.509) (255,-3.818) (256,-4.586) (257,-4.861) (258,-4.928) (259,-4.335) (260,-5.291) (261,-5.340) (262,-4.146) (263,-2.941) (264,-2.635) (265,-3.491) (266,-4.589) (267,-5.066) (268,-4.270) (269,-3.463) (270,-4.251) (271,-4.285) (272,-3.202) (273,-3.298) (274,-2.260) (275,0.179) (276,-2.186) (277,-2.623) (278,-1.571) (279,-2.081) (280,-0.929) (281,-1.008) (282,-0.565) (283,0.618) (284,0.355) (285,1.186) (286,0.492) (287,0.992) (288,1.481) (289,0.798) (290,1.383) (291,1.678) (292,0.913) (293,1.140) (294,1.777) (295,1.208) (296,0.764) (297,1.136) (298,1.428) (299,0.980) (300,0.486) (301,1.202) (302,1.216) (303,0.295) (304,0.975) (305,1.181) (306,0.265) (307,0.893) (308,1.005) (309,0.286) (310,0.983) (311,0.315) (312,0.871) (313,0.942) (314,0.192) (315,0.981) (316,0.206) (317,0.938) (318,0.120) (319,0.980) (320,0.161) (321,0.877) (322,0.167) (323,0.934) (324,0.172) (325,0.851) (326,0.202) (327,0.908) (328,0.104) (329,0.365) (330,0.722) (331,0.138) (332,0.781) (333,0.190) (334,0.284) (335,1.790) (336,0.841) (337,0.238) (338,0.556) (339,-0.476) (340,-0.373) (341,0.353) (342,-0.418) (343,-0.110) (344,0.799) (345,0.087) (346,-0.080) (347,0.387) (348,-0.561) (349,-1.100) (350,-0.724) (351,-2.956) (352,-2.608) (353,-3.980) (354,-3.321) (355,-4.365) (356,-1.916) (357,-2.143) (358,-1.313) (359,-1.952) (360,-2.297) (361,-3.126) (362,-3.026) (363,-3.037) (364,-2.911) (365,-3.460) (366,-3.428) (367,-4.095) (368,-2.814) (369,-3.102) (370,-3.880) (371,-3.750) (372,-3.741) (373,-3.888) (374,-2.809) (375,-2.239) (376,-1.383) (377,-0.531) (378,0.416) (379,0.428) (380,-0.104) (381,-0.099) (382,0.310) (383,0.105) (384,-0.547) (385,-0.178) (386,0.676) (387,0.760) (388,0.219) (389,0.966) (390,1.128) (391,0.492) (392,0.719) (393,1.091) (394,0.292) (395,0.930) (396,0.702) (397,0.447) (398,0.993) (399,0.101) (400,0.940) (401,0.200) (402,0.887) (403,0.105) (404,0.929) (405,0.178) (406,0.855) (407,0.145) (408,0.441) (409,0.607) (410,0.097) (411,0.875) (412,0.184) (413,0.097) (414,0.809) (415,0.202) (416,0.150) (417,0.772) (418,0.185) (419,0.200) (420,0.778) (421,0.123) (422,1.426) (423,0.556) (424,0.336) (425,0.172) (426,0.200) (427,0.739) (428,0.110) (429,0.201) (430,0.169) (431,0.098) (432,0.388) (433,0.495) (434,-0.116) (435,-0.406) (436,-0.669) (437,-1.075) (438,-0.997) (439,-2.035) (440,-2.421) (441,-2.891) (442,-3.278) (443,-2.842) (444,-3.795) (445,-4.151) (446,-3.619) (447,-4.557) (448,-3.463) (449,-2.815) (450,-2.034) (451,-5.419) (452,-4.320) (453,-4.585) (454,-4.848) (455,-4.602) (456,-3.133) (457,-5.419) (458,-4.394) (459,-3.939) (460,-4.896) (461,-4.233) (462,-3.829) (463,-2.172) (464,-1.832) (465,-2.410) (466,-1.802) (467,-3.657) (468,-2.791) (469,-0.448) (470,-1.623) (471,-2.217) (472,-0.751) (473,-0.865) (474,0.141) (475,0.252) (476,-0.320) (477,0.619) (478,1.231) (479,0.537) (480,0.812) (481,1.394) (482,0.945) (483,0.878) (484,1.161) (485,1.505) (486,1.068) (487,3.203) (488,1.146) (489,1.324) (490,0.791) (491,0.564) (492,1.157) (493,0.996) (494,0.278) (495,1.043) (496,0.949) (497,0.318) (498,1.007) (499,0.218) (500,0.976) (501,0.818) (502,0.223) (503,0.988) (504,0.170) (505,0.915) (506,0.148) (507,0.974) (508,0.133) (509,0.864) (510,0.201) (511,0.838) (512,0.220) (513,0.139) (514,0.928) (515,0.143) (516,0.820) (517,0.190) (518,0.325) (519,0.687) (520,0.131) (521,0.645) (522,0.421) (523,0.098) (524,0.180) (525,0.867) (526,0.102) (527,0.127) (528,0.593) (529,0.340) (530,-0.094) (531,-0.065) (532,0.508) (533,-0.449) (534,-0.474) (535,-0.839) (536,-0.782) (537,-1.595) (538,-2.524) (539,-2.702) (540,-4.100) (541,-3.906) (542,-3.559) (543,-4.414) (544,-1.940) (545,-4.648) (546,-2.899) (547,-3.801) (548,-3.790) (549,-3.579) (550,-4.046) (551,-4.040) (552,-5.013) (553,-3.397) (554,-4.888) (555,-5.373) (556,-4.333) (557,-3.482) (558,-4.702) (559,-2.907) (560,-3.177) (561,-3.646) (562,-3.875) (563,-3.121) (564,-2.519) (565,-2.660) (566,-1.475) (567,-1.329) (568,-2.022) (569,-0.534) (570,-0.747) (571,0.323) (572,-0.288) (573,1.005) (574,0.559) (575,1.074) (576,1.177) (577,0.873) (578,1.862) (579,0.997) (580,1.032) (581,4.314) (582,1.292) (583,0.879) (584,1.109) (585,1.493) (586,1.145) (587,0.699) (588,0.971) (589,1.252) (590,0.769) (591,0.589) (592,1.183) (593,1.030) (594,0.238) (595,1.143) (596,1.062) (597,0.110) (598,1.096) (599,0.795) (600,0.432) (601,0.982) (602,0.208) (603,1.029) (604,0.530) (605,0.600) (606,1.031) (607,0.101) (608,0.972) (609,0.207) (610,1.017) (611,0.098) (612,0.915) (613,0.229) (614,0.861) (615,0.123) (616,0.987) (617,0.170) (618,0.835) (619,0.171) (620,0.423) (621,0.631) (622,0.117) (623,0.876) (624,0.170) (625,0.065) (626,0.668) (627,-0.233) (628,-0.641) (629,-0.291) (630,-1.347) (631,-0.972) (632,-1.511) (633,-2.154) (634,-1.556) (635,-2.736) (636,-3.110) (637,-2.679) (638,-3.596) (639,-3.353) (640,-3.730) (641,-4.363) (642,-3.906) (643,-4.692) (644,-4.464) (645,-4.196) (646,-4.680) (647,-4.932) (648,-4.246) (649,-4.118) (650,-4.788) (651,-4.662) (652,-3.906) (653,-3.826) (654,-4.286) (655,-3.779) (656,-3.211) (657,-3.571) (658,-3.573) (659,-2.828) (660,-2.587) (661,-2.519) (662,-1.586) (663,-1.568) (664,-1.617) (665,-0.192) (666,-0.705) (667,0.550) (668,-0.068) (669,0.680) (670,0.843) (671,0.493) (672,1.712)};
\end{axis}
\end{tikzpicture}
\end{minipage}

\vspace{0.5em}

\begin{minipage}[t]{0.48\textwidth}
\centering
\begin{tikzpicture}
\begin{axis}[
width=\textwidth,
height=0.24\textheight,
title={Home C: high evening peaks},
xmin=1, xmax=672,
ymin=-6, ymax=18,
ytick={0,5,10,15},
xtick={1,97,193,289,385,481,577},
xticklabels={Aug 1,Aug 2,Aug 3,Aug 4,Aug 5,Aug 6,Aug 7},
xticklabel style={font=\scriptsize,rotate=45,anchor=east},
xlabel={15-minute interval in Base week},
ylabel={Net load (kW)},
grid=both,
major grid style={gray!20},
minor grid style={gray!10}
]
\addplot+[mark=none, line width=0.7pt] coordinates {(1,4.409) (2,4.333) (3,4.281) (4,4.300) (5,4.501) (6,4.234) (7,4.654) (8,4.664) (9,3.955) (10,4.154) (11,4.673) (12,3.268) (13,4.660) (14,3.417) (15,4.148) (16,3.398) (17,3.605) (18,2.855) (19,2.052) (20,3.720) (21,2.617) (22,1.513) (23,3.116) (24,1.927) (25,2.458) (26,2.913) (27,2.355) (28,2.021) (29,2.432) (30,2.863) (31,2.650) (32,1.955) (33,2.328) (34,2.985) (35,2.155) (36,1.822) (37,3.044) (38,2.165) (39,1.776) (40,2.192) (41,2.560) (42,2.741) (43,1.067) (44,2.675) (45,1.609) (46,1.813) (47,2.453) (48,0.771) (49,2.748) (50,0.765) (51,2.542) (52,1.978) (53,3.933) (54,2.392) (55,2.776) (56,3.248) (57,3.499) (58,3.254) (59,2.816) (60,2.639) (61,3.416) (62,4.045) (63,3.285) (64,3.592) (65,3.522) (66,2.961) (67,3.505) (68,4.288) (69,4.366) (70,2.388) (71,4.440) (72,5.070) (73,4.443) (74,4.286) (75,4.384) (76,4.948) (77,4.664) (78,5.807) (79,5.792) (80,5.326) (81,5.297) (82,5.209) (83,5.210) (84,4.996) (85,5.121) (86,5.479) (87,5.393) (88,5.259) (89,4.728) (90,5.591) (91,4.636) (92,4.540) (93,4.688) (94,4.386) (95,4.264) (96,4.172) (97,4.088) (98,4.046) (99,4.072) (100,4.111) (101,4.123) (102,4.297) (103,4.651) (104,4.852) (105,4.079) (106,4.487) (107,3.142) (108,5.444) (109,7.069) (110,6.549) (111,5.746) (112,4.099) (113,3.459) (114,3.068) (115,6.168) (116,2.455) (117,2.558) (118,2.451) (119,2.249) (120,2.573) (121,2.585) (122,2.652) (123,2.425) (124,2.382) (125,2.528) (126,2.690) (127,2.452) (128,2.137) (129,2.139) (130,2.563) (131,2.322) (132,2.371) (133,2.438) (134,2.347) (135,2.341) (136,2.410) (137,2.259) (138,2.388) (139,2.308) (140,2.052) (141,2.228) (142,2.141) (143,2.298) (144,2.237) (145,2.416) (146,2.300) (147,1.826) (148,2.199) (149,2.634) (150,2.604) (151,2.693) (152,2.691) (153,2.867) (154,2.818) (155,4.312) (156,4.554) (157,3.217) (158,4.161) (159,3.521) (160,4.372) (161,3.373) (162,4.602) (163,4.627) (164,4.108) (165,4.051) (166,4.810) (167,4.629) (168,4.770) (169,4.648) (170,4.270) (171,4.336) (172,4.469) (173,5.574) (174,4.643) (175,4.301) (176,4.146) (177,4.059) (178,4.049) (179,3.967) (180,3.154) (181,2.457) (182,4.613) (183,4.280) (184,4.292) (185,2.795) (186,3.837) (187,3.453) (188,3.683) (189,3.114) (190,3.782) (191,4.145) (192,2.714) (193,3.244) (194,3.312) (195,3.578) (196,5.593) (197,4.524) (198,3.971) (199,3.537) (200,3.019) (201,2.859) (202,3.729) (203,3.596) (204,3.103) (205,3.287) (206,2.953) (207,2.902) (208,2.441) (209,2.332) (210,2.272) (211,2.403) (212,2.614) (213,2.561) (214,2.221) (215,2.421) (216,2.127) (217,2.066) (218,2.056) (219,1.894) (220,1.748) (221,1.318) (222,1.667) (223,1.967) (224,2.231) (225,1.813) (226,1.517) (227,1.693) (228,1.987) (229,1.798) (230,1.133) (231,1.959) (232,1.741) (233,0.998) (234,2.008) (235,1.268) (236,1.489) (237,1.929) (238,0.755) (239,1.888) (240,0.608) (241,1.868) (242,0.789) (243,1.997) (244,1.968) (245,2.657) (246,2.781) (247,2.570) (248,2.722) (249,2.193) (250,2.004) (251,2.276) (252,2.736) (253,2.642) (254,2.594) (255,2.627) (256,2.855) (257,9.296) (258,13.045) (259,14.019) (260,13.693) (261,13.118) (262,14.009) (263,12.829) (264,17.828) (265,16.105) (266,13.198) (267,4.354) (268,2.813) (269,4.501) (270,4.499) (271,4.438) (272,4.400) (273,4.413) (274,4.642) (275,4.593) (276,4.304) (277,4.475) (278,4.408) (279,4.440) (280,4.379) (281,4.411) (282,4.391) (283,4.270) (284,4.301) (285,4.303) (286,4.293) (287,4.289) (288,4.299) (289,4.304) (290,4.477) (291,4.387) (292,4.670) (293,8.225) (294,8.065) (295,7.260) (296,5.505) (297,4.384) (298,3.205) (299,5.025) (300,3.680) (301,3.962) (302,3.917) (303,3.488) (304,3.074) (305,3.294) (306,3.487) (307,2.879) (308,2.701) (309,2.598) (310,2.556) (311,2.580) (312,2.687) (313,2.529) (314,2.418) (315,2.439) (316,2.405) (317,2.434) (318,2.206) (319,2.177) (320,2.174) (321,2.207) (322,2.305) (323,2.468) (324,2.331) (325,2.178) (326,2.288) (327,2.254) (328,2.417) (329,2.286) (330,2.162) (331,2.034) (332,2.275) (333,2.047) (334,2.091) (335,1.845) (336,1.300) (337,1.635) (338,2.164) (339,2.221) (340,1.865) (341,3.069) (342,2.695) (343,2.784) (344,2.783) (345,2.768) (346,3.233) (347,3.148) (348,2.655) (349,2.937) (350,2.877) (351,2.988) (352,2.789) (353,2.714) (354,2.765) (355,2.460) (356,2.945) (357,12.681) (358,12.893) (359,13.318) (360,13.249) (361,13.983) (362,12.528) (363,12.511) (364,13.109) (365,13.887) (366,13.071) (367,13.830) (368,14.261) (369,13.027) (370,14.603) (371,14.634) (372,12.240) (373,14.559) (374,14.599) (375,14.583) (376,14.440) (377,12.616) (378,14.446) (379,14.467) (380,13.859) (381,6.896) (382,5.009) (383,4.434) (384,4.435) (385,4.782) (386,4.903) (387,3.357) (388,4.316) (389,4.330) (390,2.472) (391,4.643) (392,4.435) (393,1.767) (394,3.916) (395,3.460) (396,3.357) (397,3.738) (398,4.410) (399,4.134) (400,2.747) (401,3.602) (402,2.232) (403,2.220) (404,2.963) (405,2.412) (406,1.803) (407,1.904) (408,2.471) (409,2.824) (410,1.852) (411,1.878) (412,2.466) (413,2.837) (414,1.824) (415,1.990) (416,2.614) (417,2.481) (418,1.684) (419,1.631) (420,2.515) (421,1.530) (422,1.825) (423,2.436) (424,0.831) (425,2.328) (426,1.478) (427,1.676) (428,1.890) (429,1.113) (430,1.875) (431,1.067) (432,2.072) (433,0.905) (434,2.360) (435,0.821) (436,2.917) (437,1.901) (438,2.465) (439,1.226) (440,2.825) (441,2.534) (442,1.223) (443,2.683) (444,1.872) (445,1.387) (446,2.741) (447,2.340) (448,2.831) (449,2.351) (450,3.120) (451,2.288) (452,1.920) (453,3.392) (454,2.437) (455,2.242) (456,2.768) (457,3.509) (458,2.136) (459,4.088) (460,2.114) (461,4.152) (462,4.193) (463,4.241) (464,2.046) (465,4.299) (466,4.370) (467,4.245) (468,5.029) (469,4.517) (470,4.481) (471,4.490) (472,4.436) (473,4.249) (474,4.420) (475,4.562) (476,4.413) (477,4.399) (478,4.315) (479,4.338) (480,4.202) (481,4.157) (482,4.112) (483,2.448) (484,3.884) (485,1.636) (486,3.385) (487,3.222) (488,2.369) (489,3.362) (490,3.756) (491,4.349) (492,5.060) (493,4.910) (494,4.953) (495,3.900) (496,4.833) (497,3.225) (498,4.008) (499,3.338) (500,3.975) (501,3.944) (502,3.919) (503,2.898) (504,2.729) (505,2.643) (506,2.852) (507,3.010) (508,2.864) (509,2.736) (510,2.249) (511,2.316) (512,2.347) (513,2.164) (514,2.199) (515,2.160) (516,2.246) (517,2.219) (518,2.114) (519,2.105) (520,2.355) (521,2.281) (522,2.341) (523,2.297) (524,2.025) (525,2.124) (526,1.908) (527,2.085) (528,2.162) (529,2.505) (530,3.314) (531,6.146) (532,4.941) (533,2.371) (534,0.903) (535,0.905) (536,1.011) (537,2.464) (538,0.909) (539,2.573) (540,2.160) (541,1.218) (542,2.725) (543,2.746) (544,1.720) (545,1.616) (546,2.782) (547,2.723) (548,1.343) (549,3.034) (550,2.226) (551,1.526) (552,3.647) (553,1.941) (554,4.060) (555,2.271) (556,4.391) (557,4.430) (558,4.507) (559,4.524) (560,4.403) (561,4.471) (562,4.411) (563,4.303) (564,4.346) (565,4.353) (566,4.434) (567,4.392) (568,4.428) (569,4.441) (570,4.472) (571,4.434) (572,4.456) (573,4.438) (574,4.422) (575,4.279) (576,4.234) (577,4.227) (578,4.091) (579,4.036) (580,4.127) (581,5.710) (582,14.513) (583,14.883) (584,13.330) (585,13.804) (586,13.079) (587,12.953) (588,13.834) (589,14.094) (590,12.282) (591,13.394) (592,12.773) (593,14.303) (594,14.601) (595,14.068) (596,14.077) (597,13.856) (598,13.146) (599,12.236) (600,14.158) (601,12.934) (602,12.460) (603,14.017) (604,12.478) (605,12.554) (606,13.118) (607,11.588) (608,4.278) (609,4.390) (610,3.171) (611,2.014) (612,2.116) (613,2.957) (614,2.756) (615,1.942) (616,1.909) (617,2.864) (618,2.904) (619,1.906) (620,1.574) (621,2.660) (622,2.641) (623,1.748) (624,2.252) (625,2.953) (626,3.958) (627,2.005) (628,2.130) (629,2.719) (630,2.827) (631,2.857) (632,1.531) (633,2.117) (634,3.081) (635,2.329) (636,1.844) (637,2.680) (638,3.094) (639,1.966) (640,2.144) (641,3.123) (642,2.235) (643,2.046) (644,3.225) (645,2.313) (646,4.580) (647,4.251) (648,3.349) (649,2.699) (650,4.108) (651,4.112) (652,4.118) (653,4.474) (654,4.570) (655,5.679) (656,8.087) (657,7.654) (658,7.425) (659,7.589) (660,7.351) (661,6.208) (662,4.847) (663,4.688) (664,4.675) (665,5.106) (666,4.561) (667,4.605) (668,4.740) (669,4.708) (670,4.688) (671,4.585) (672,4.498)};
\end{axis}
\end{tikzpicture}
\end{minipage}
\hfill
\begin{minipage}[t]{0.48\textwidth}
\centering
\begin{tikzpicture}
\begin{axis}[
width=\textwidth,
height=0.24\textheight,
title={Home D: low-load, flatter},
xmin=1, xmax=672,
ymin=-6, ymax=18,
ytick={0,5,10,15},
xtick={1,97,193,289,385,481,577},
xticklabels={Aug 1,Aug 2,Aug 3,Aug 4,Aug 5,Aug 6,Aug 7},
xticklabel style={font=\scriptsize,rotate=45,anchor=east},
xlabel={15-minute interval in Base week},
grid=both,
major grid style={gray!20},
minor grid style={gray!10}
]
\addplot+[mark=none, line width=0.7pt] coordinates {(1,0.424) (2,0.419) (3,0.379) (4,0.346) (5,0.327) (6,0.343) (7,0.436) (8,0.479) (9,0.548) (10,0.538) (11,0.356) (12,0.408) (13,0.428) (14,0.382) (15,0.347) (16,0.309) (17,0.434) (18,0.321) (19,0.341) (20,0.254) (21,0.269) (22,0.374) (23,0.398) (24,0.313) (25,0.225) (26,0.280) (27,0.266) (28,0.216) (29,0.279) (30,0.340) (31,0.280) (32,0.175) (33,0.188) (34,0.247) (35,0.270) (36,0.232) (37,0.302) (38,0.247) (39,0.145) (40,0.191) (41,0.228) (42,0.163) (43,0.314) (44,0.339) (45,0.243) (46,0.319) (47,0.213) (48,0.192) (49,0.309) (50,0.197) (51,0.244) (52,0.151) (53,0.181) (54,0.215) (55,0.182) (56,0.317) (57,0.277) (58,0.150) (59,0.242) (60,0.247) (61,0.239) (62,0.486) (63,0.523) (64,0.387) (65,0.236) (66,0.274) (67,0.379) (68,0.341) (69,0.508) (70,0.496) (71,0.454) (72,0.349) (73,0.374) (74,0.351) (75,0.410) (76,0.418) (77,0.390) (78,0.328) (79,0.327) (80,0.328) (81,0.334) (82,0.422) (83,0.419) (84,0.385) (85,0.328) (86,0.325) (87,0.325) (88,0.354) (89,0.419) (90,0.416) (91,0.365) (92,0.327) (93,0.325) (94,0.327) (95,0.378) (96,0.420) (97,0.417) (98,0.348) (99,0.245) (100,0.250) (101,0.457) (102,0.366) (103,0.282) (104,0.269) (105,0.257) (106,0.211) (107,0.159) (108,0.431) (109,0.669) (110,0.636) (111,0.558) (112,0.341) (113,0.440) (114,0.340) (115,0.532) (116,0.583) (117,0.636) (118,0.536) (119,0.559) (120,0.527) (121,0.484) (122,0.603) (123,0.678) (124,0.703) (125,0.485) (126,0.576) (127,0.573) (128,0.611) (129,0.654) (130,0.375) (131,0.256) (132,0.293) (133,0.187) (134,0.277) (135,0.246) (136,0.377) (137,0.598) (138,0.543) (139,0.294) (140,0.201) (141,0.182) (142,0.269) (143,0.241) (144,0.189) (145,0.152) (146,0.114) (147,0.261) (148,0.206) (149,0.309) (150,0.112) (151,0.216) (152,0.109) (153,0.237) (154,0.203) (155,0.318) (156,0.261) (157,0.235) (158,0.242) (159,0.195) (160,0.241) (161,0.368) (162,0.419) (163,0.281) (164,0.319) (165,0.379) (166,0.377) (167,0.467) (168,0.488) (169,0.468) (170,0.483) (171,0.568) (172,0.574) (173,0.618) (174,0.666) (175,0.661) (176,0.615) (177,0.586) (178,0.577) (179,0.504) (180,0.574) (181,0.650) (182,0.554) (183,0.457) (184,0.493) (185,0.490) (186,0.411) (187,0.474) (188,0.564) (189,0.509) (190,0.410) (191,0.481) (192,0.393) (193,0.414) (194,0.598) (195,0.685) (196,0.461) (197,0.221) (198,0.102) (199,0.208) (200,0.136) (201,0.302) (202,0.239) (203,0.557) (204,0.878) (205,0.635) (206,0.574) (207,0.580) (208,0.457) (209,0.423) (210,0.466) (211,0.460) (212,0.321) (213,0.386) (214,0.389) (215,0.346) (216,0.180) (217,0.279) (218,0.347) (219,0.169) (220,0.213) (221,0.328) (222,0.163) (223,0.230) (224,0.172) (225,0.335) (226,0.183) (227,0.183) (228,0.277) (229,0.213) (230,0.209) (231,0.198) (232,0.289) (233,0.153) (234,0.153) (235,0.250) (236,0.192) (237,0.220) (238,0.113) (239,0.315) (240,0.167) (241,0.159) (242,0.211) (243,0.246) (244,0.460) (245,0.351) (246,0.452) (247,0.231) (248,0.227) (249,0.117) (250,0.300) (251,0.307) (252,0.138) (253,0.246) (254,0.346) (255,0.266) (256,0.212) (257,0.462) (258,0.926) (259,0.815) (260,0.824) (261,0.959) (262,0.734) (263,0.631) (264,0.672) (265,0.769) (266,0.821) (267,0.794) (268,0.829) (269,0.874) (270,0.827) (271,0.771) (272,0.846) (273,0.939) (274,0.889) (275,0.889) (276,0.970) (277,0.933) (278,0.884) (279,0.809) (280,0.791) (281,0.838) (282,0.910) (283,0.864) (284,0.613) (285,0.601) (286,0.419) (287,0.526) (288,0.730) (289,0.775) (290,0.662) (291,0.446) (292,0.512) (293,0.680) (294,0.750) (295,0.805) (296,0.907) (297,0.963) (298,0.841) (299,0.827) (300,0.921) (301,1.008) (302,1.067) (303,1.100) (304,1.282) (305,1.295) (306,1.194) (307,1.225) (308,1.298) (309,1.222) (310,1.147) (311,0.802) (312,0.720) (313,0.743) (314,0.594) (315,0.572) (316,0.676) (317,0.584) (318,0.494) (319,0.613) (320,0.711) (321,0.737) (322,0.669) (323,0.619) (324,0.608) (325,0.550) (326,0.671) (327,0.653) (328,0.362) (329,0.242) (330,0.336) (331,0.285) (332,0.290) (333,0.278) (334,0.159) (335,0.334) (336,0.304) (337,0.251) (338,0.272) (339,0.161) (340,0.351) (341,0.292) (342,0.240) (343,0.278) (344,0.170) (345,0.373) (346,0.278) (347,0.238) (348,0.277) (349,0.217) (350,0.468) (351,0.276) (352,0.266) (353,0.271) (354,0.253) (355,0.415) (356,0.301) (357,0.274) (358,0.374) (359,0.390) (360,0.400) (361,0.359) (362,0.360) (363,0.279) (364,0.351) (365,0.395) (366,0.379) (367,0.279) (368,0.268) (369,0.386) (370,0.435) (371,0.280) (372,0.288) (373,0.372) (374,0.379) (375,0.409) (376,0.343) (377,0.287) (378,0.347) (379,0.439) (380,0.611) (381,0.642) (382,0.660) (383,0.602) (384,0.538) (385,0.624) (386,0.431) (387,0.444) (388,0.672) (389,0.846) (390,0.840) (391,0.796) (392,0.818) (393,0.723) (394,0.809) (395,0.837) (396,0.795) (397,0.436) (398,0.411) (399,0.775) (400,0.455) (401,0.398) (402,0.385) (403,0.376) (404,0.441) (405,0.733) (406,0.765) (407,0.671) (408,0.473) (409,0.470) (410,0.471) (411,0.343) (412,0.320) (413,0.357) (414,0.287) (415,0.411) (416,0.415) (417,0.222) (418,0.333) (419,0.395) (420,0.629) (421,0.513) (422,0.535) (423,0.493) (424,0.535) (425,0.592) (426,0.543) (427,0.561) (428,0.456) (429,0.484) (430,0.608) (431,0.520) (432,0.515) (433,0.410) (434,0.423) (435,0.667) (436,0.774) (437,0.857) (438,0.772) (439,0.841) (440,0.929) (441,0.846) (442,0.836) (443,0.849) (444,0.622) (445,0.658) (446,0.637) (447,0.443) (448,0.456) (449,0.518) (450,0.524) (451,0.615) (452,0.695) (453,0.658) (454,0.743) (455,0.751) (456,0.807) (457,0.730) (458,0.619) (459,0.589) (460,0.576) (461,0.625) (462,0.593) (463,0.527) (464,0.537) (465,0.567) (466,0.628) (467,0.596) (468,0.534) (469,0.532) (470,0.569) (471,0.626) (472,0.593) (473,0.537) (474,0.545) (475,0.573) (476,0.612) (477,0.595) (478,0.519) (479,0.456) (480,0.533) (481,0.611) (482,0.529) (483,0.510) (484,0.436) (485,0.508) (486,0.520) (487,0.580) (488,0.442) (489,0.408) (490,0.433) (491,0.549) (492,0.489) (493,0.382) (494,0.411) (495,0.506) (496,0.494) (497,0.484) (498,0.667) (499,0.715) (500,0.491) (501,0.518) (502,0.617) (503,0.552) (504,0.470) (505,0.469) (506,0.597) (507,0.553) (508,0.443) (509,0.419) (510,0.464) (511,0.476) (512,0.369) (513,0.331) (514,0.495) (515,0.523) (516,0.465) (517,0.551) (518,0.418) (519,0.229) (520,0.353) (521,0.474) (522,0.577) (523,0.529) (524,0.185) (525,0.196) (526,0.187) (527,0.266) (528,0.254) (529,0.209) (530,0.122) (531,0.224) (532,0.197) (533,0.301) (534,0.177) (535,0.144) (536,0.205) (537,0.159) (538,0.376) (539,0.283) (540,0.220) (541,0.252) (542,0.194) (543,0.314) (544,0.361) (545,0.394) (546,0.600) (547,0.547) (548,0.433) (549,0.420) (550,0.462) (551,0.471) (552,0.441) (553,0.365) (554,0.369) (555,0.431) (556,0.411) (557,0.338) (558,0.346) (559,0.401) (560,0.435) (561,0.394) (562,0.344) (563,0.344) (564,0.425) (565,0.433) (566,0.298) (567,0.343) (568,0.350) (569,0.437) (570,0.444) (571,0.361) (572,0.347) (573,0.345) (574,0.414) (575,0.432) (576,0.291) (577,0.341) (578,0.339) (579,0.468) (580,0.560) (581,0.562) (582,0.460) (583,0.527) (584,0.517) (585,0.535) (586,0.524) (587,0.320) (588,0.208) (589,0.260) (590,0.359) (591,0.264) (592,0.198) (593,0.228) (594,0.330) (595,0.362) (596,0.310) (597,0.297) (598,0.352) (599,0.339) (600,0.354) (601,0.385) (602,0.306) (603,0.205) (604,0.272) (605,0.392) (606,0.293) (607,0.160) (608,0.250) (609,0.305) (610,0.274) (611,0.310) (612,0.187) (613,0.178) (614,0.283) (615,0.224) (616,0.309) (617,0.166) (618,0.343) (619,0.369) (620,0.339) (621,0.204) (622,0.199) (623,0.200) (624,0.196) (625,0.321) (626,0.185) (627,0.233) (628,0.121) (629,0.264) (630,0.216) (631,0.300) (632,0.155) (633,0.205) (634,0.254) (635,0.244) (636,0.395) (637,0.285) (638,0.389) (639,0.427) (640,0.469) (641,0.356) (642,0.265) (643,0.338) (644,0.372) (645,0.482) (646,0.411) (647,0.390) (648,0.411) (649,0.407) (650,0.461) (651,0.413) (652,0.344) (653,0.349) (654,0.359) (655,0.431) (656,0.398) (657,0.337) (658,0.335) (659,0.358) (660,0.426) (661,0.395) (662,0.334) (663,0.334) (664,0.359) (665,0.426) (666,0.396) (667,0.334) (668,0.334) (669,0.363) (670,0.427) (671,0.391) (672,0.336)};
\end{axis}
\end{tikzpicture}
\end{minipage}
\caption{Base-week net load profiles for four representative homes. Net load is defined as household load minus contemporaneous solar generation, $N_{g,\tau}=L_{g,\tau}-S_{g,\tau}$. Each panel plots the full week from August~1 through August~7, 2025, at 15-minute resolution, with x-axis labels marking the start of each day. The selected homes illustrate four qualitatively different patterns in the data: a typical non-solar home, a solar-heavy home with midday negative net load, a home with pronounced evening peaks, and a lower-load flatter home.}
\label{fig:selected-home-net-loads}
\end{figure}
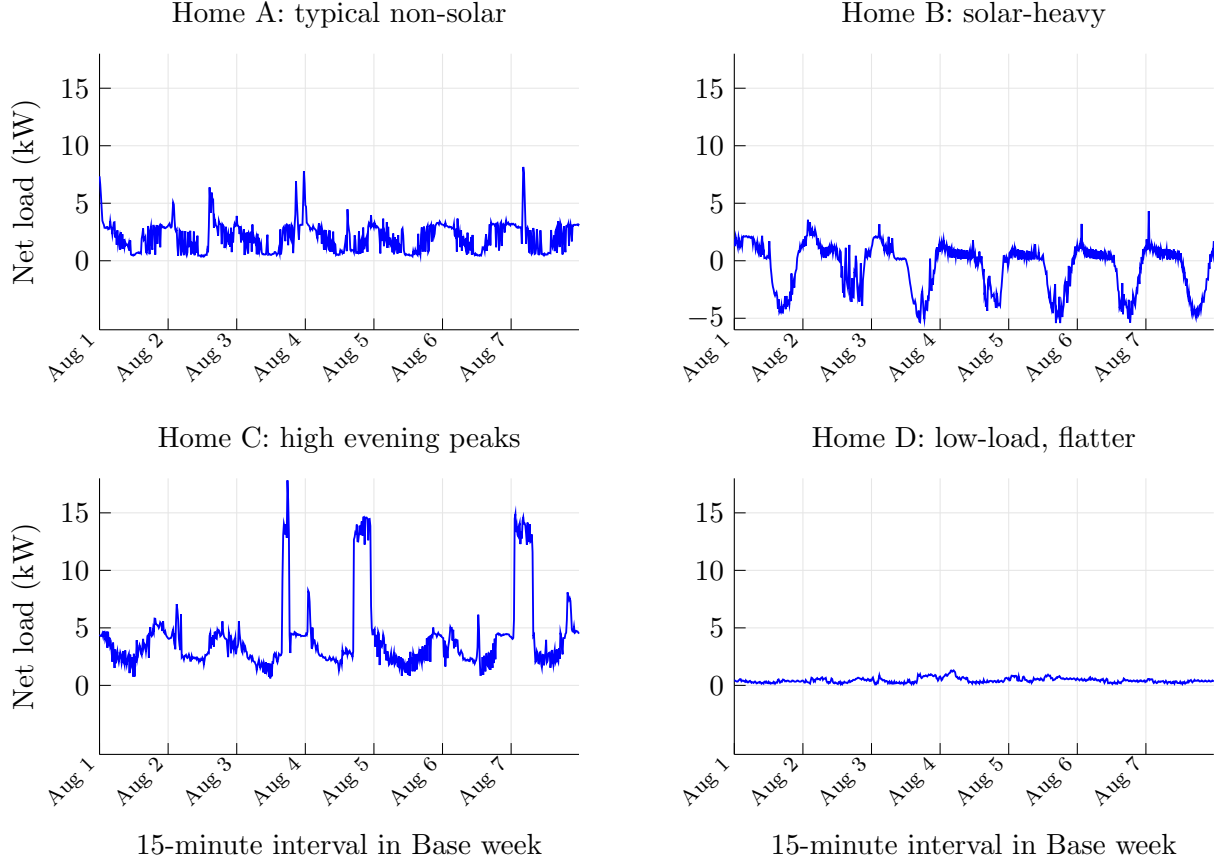

Figure~\ref{fig:ercot-zones} places the sample geographically inside ERCOT. The Base homes are located in the South load zone, so the wholesale series used throughout the paper is the ERCOT real-time price at settlement point \texttt{LZ\_SOUTH}. Figure~\ref{fig:selected-home-net-loads} complements that market-level view with household-level heterogeneity. It plots 15-minute net load profiles, defined as $N_{g,\tau}=L_{g,\tau}-S_{g,\tau}$, for four representative homes over the empirical week. The homes are chosen to illustrate a typical non-solar home, a solar-heavy home with midday negative net load, a home with pronounced evening peaks, and a low-load flatter home.

The economic inputs combine ERCOT real-time prices with a benchmark retail tariff motivated by Base's publicly described contract terms. The tariff uses a retail energy charge $p^{\mathrm{ret}}$, an import-side delivery charge $C_{\mathrm{TDSP}}$, and a solar credit $\beta$ for customer-owned solar that is not immediately self-consumed. Throughout the empirical analysis we set
\begin{equation}
 p^{\mathrm{ret}}=0.09\text{ USD/kWh},
 \qquad
 C_{\mathrm{TDSP}}=0.05\text{ USD/kWh},
 \qquad
 \beta=0.04\text{ USD/kWh}.
 \label{eq:baseline-tariff-parameters}
\end{equation}
These tariff parameters enter both the MPC objective and the ex post firm-margin accounting. Fixed subscription revenue is treated separately: because it does not vary with dispatch over the comparisons studied here, it is omitted from the MPC objective and added only to reported total firm margin.
Although retail energy revenue is fixed for a realized load path, we keep $p^{\mathrm{ret}}L$ in the per-period margin so the MPC objective and ex post dispatch margin use the same accounting convention; fixed weekly subscription revenue is added only when reporting total firm margin.

\section{Control Model}\label{sec:model}

Standalone service and pooled service are distinct operational problems. They share the same data inputs, reserve logic, and MPC timing, but they differ in what actions are feasible and in how value is created.
\begin{enumerate}
\item \textbf{Standalone operation:} each home with its own battery, modeled through a dispatch LP with the routing variables needed to represent the provider's cash flows.
\item \textbf{Pooling:} many homes in a fixed pool, with one battery state and one backup requirement per home, and with operational coordination through internal energy sharing across homes.
\end{enumerate}

We model pooling as provider-controlled coordination of a managed fleet rather than as aggregation into a single physical battery. In the pooled LP, the sharing variables allow the provider to net contemporaneous imports, exports, charging, and discharging across enrolled homes for dispatch and accounting purposes. Each household still maintains its own state of charge and backup reserve floor.

\subsection{Common Setup and Timing}\label{subsec:common-setup}

\begin{table}[H]
\centering
\small
\begin{tabular}{@{}p{0.24\textwidth}p{0.54\textwidth}p{0.10\textwidth}@{}}
\toprule
Symbol & Meaning & Units \\
\midrule
$\mathcal{G}$ & set of homes in the one fixed pool & -- \\
$g$ & generic home index & -- \\
$\tau$ & generic calendar interval on the 15-minute grid & -- \\
$\Delta$ & length of one control interval & hours \\
$L_{g,\tau}$ & average household load of home $g$ during interval $\tau$ & kW \\
$S_{g,\tau}$ & average solar generation of home $g$ during interval $\tau$ & kW \\
$N_{g,\tau}$ & average net load of home $g$ during interval $\tau$, defined by $N_{g,\tau}:=L_{g,\tau}-S_{g,\tau}$ & kW \\
$E_g^{\max}$ & usable battery energy capacity of home $g$ & kWh \\
$P^{\max}_{g,\mathrm{ch}}$ & home charging power limit & kW \\
$P^{\max}_{g,\mathrm{dis}}$ & home discharging power limit & kW \\
$\eta_{g,\mathrm{ch}}$ & home charging efficiency & -- \\
$\eta_{g,\mathrm{dis}}$ & home discharging efficiency & -- \\
$\lambda_{\tau}$ & realized wholesale price in calendar interval $\tau$ & USD/kWh \\

$p^{\mathrm{ret}}$ & retail energy charge per kWh of household energy service & USD/kWh \\
$C_{\mathrm{TDSP}}$ & import-side delivery charge / TDSP wedge applied per kWh of metered grid import & USD/kWh \\
$\beta$ & solar credit per kWh of customer-owned solar not directly self-consumed & USD/kWh \\
\bottomrule
\end{tabular}
\caption{Core physical and economic notation for the control model.}
\label{tab:shared-notation}
\end{table}

Let $\mathcal{G}$ denote the set of homes in the fixed pool considered in the pooling section. The decision interval is 15 minutes, so one period has length
\[
\Delta = 0.25 \text{ hours}.
\]
At any decision epoch $t$, the controller optimizes over a horizon of $H$ future control periods, indexed by
\[
h=0,1,\ldots,H-1,
\]
with baseline value $H=96$ for a 24-hour horizon. Throughout the paper, both the standalone and pooled controllers are finite-horizon \emph{model predictive control} (MPC) problems: at decision epoch $t$, the controller solves a horizon-$H$ optimization using forecasts for calendar intervals $t,t+1,\ldots,t+H-1$, implements only the first-period control, updates the realized battery state and forecasts, and then re-solves at epoch $t+1$. Thus $h$ indexes how many decision intervals ahead the controller is planning, so an object indexed by $(t,h)$ refers to the quantity for calendar interval $t+h$. State variables such as $e_{t,h+1}$ and $e_{g,t,h+1}$ denote the battery energy after the control for interval $h$ has been applied. To keep calendar-time data separate from MPC quantities indexed by decision epoch and horizon step, we use $\tau$ for a generic 15-minute interval in the shared notation below. All load, solar, charge, discharge, import, export, curtailment, and routing variables are measured as average power over a 15-minute interval (kW); multiplying by $\Delta$ converts them to interval energy (kWh).

Table~\ref{tab:shared-notation} collects the primitives used throughout the control model. The quantities $L_{g,\tau}$ and $S_{g,\tau}$ denote realized average load and solar power over calendar interval $\tau$ and come from the processed Base measurements described in the data section. The home-specific battery capacities and charge/discharge power limits are taken from Base device metadata. The efficiencies $\eta_{g,\mathrm{ch}}$ and $\eta_{g,\mathrm{dis}}$ are engineering inputs and should be read as modeling assumptions carried consistently through control, reserve construction, and ex-post reliability evaluation.

The shared economic inputs combine observed market data with the benchmark tariff described in Section~\ref{sec:data}. The realized wholesale price $\lambda_{\tau}$ comes from ERCOT real-time market data, while the control problems use forecast prices $\hat\lambda_{t,h}$. The tariff parameters $p^{\mathrm{ret}}$, $C_{\mathrm{TDSP}}$, and $\beta$ are fixed at the benchmark values in \eqref{eq:baseline-tariff-parameters}. Fixed subscription revenue is omitted from the MPC objective because it does not vary with dispatch over the comparisons studied here and therefore only shifts reported firm-margin levels.

The terminal salvage value $\lambda_{\mathrm{salv},t}$ is used only to avoid an end-of-horizon artifact in the receding-horizon controller. Unless a fixed override is supplied, the numerical implementation sets it to the median forecast avoided import cost across the current MPC horizon,
\[
\lambda_{\mathrm{salv},t}
=
\mathrm{median}_{h=0,\ldots,H-1}
\{\hat\lambda_{t,h}+C_{\mathrm{TDSP}}\}.
\]
Thus energy remaining in the battery after the final forecast interval is valued at a typical near-term avoided import cost.

In the standalone single-home problem, we suppress the home index wherever possible and keep only the time and horizon indices. In the pooling formulation, $g\in\mathcal{G}$ always denotes an individual home inside one fixed pool.

\subsection{Standalone Dispatch LP}\label{subsec:standalone-mpc}

The standalone controller solves the dispatch LP for each home, keeping the state and control variables needed for battery dynamics, reserve feasibility, and the provider's cash flows. At decision epoch $t$ the standalone controller chooses
\[
m^{\mathrm{imp}}_{t,h},\quad
u^{\mathrm{ch}}_{t,h},\quad
u^{\mathrm{dis}}_{t,h},\quad
z_{t,h},\quad
x^S_{t,h},\quad
x^B_{t,h},\quad
c_{t,h},
\qquad h=0,\ldots,H-1,
\]
together with battery energy levels
\[
e_{t,h},
\qquad h=0,\ldots,H.
\]
Here $m^{\mathrm{imp}}_{t,h}$ is metered grid import, $u^{\mathrm{ch}}_{t,h}$ and $u^{\mathrm{dis}}_{t,h}$ are total battery charge and discharge, $z_{t,h}$ is solar used to charge the battery, $x^S_{t,h}$ is solar exported to the grid, $x^B_{t,h}$ is battery discharge exported to the grid, $c_{t,h}$ is curtailed solar, and $e_{t,h}$ is internal battery energy. The formulation is ``reduced'' in the sense that it does not model all behind-the-meter electrical flows separately. It keeps exactly the flow distinctions needed for battery feasibility and tariff accounting: local solar used for charging, solar export, battery export, grid import, and curtailed solar.

Reserve targets $(r_{t,h+1})_{h=0}^{H-1}$ are treated as exogenous inputs here and are defined in Section~\ref{sec:backup-accounting}.

Given forecasts $(\hat L_{t,h},\hat S_{t,h},\hat\lambda_{t,h})_{h=0}^{H-1}$ and reserve targets $(r_{t,h+1})_{h=0}^{H-1}$, the standalone LP is
\begin{align}
\max\quad &
\sum_{h=0}^{H-1}\Pi^{\mathrm{SA}}_{t,h}
+
\lambda_{\mathrm{salv},t} e_{t,H}
\label{eq:standalone-mpc-master-obj}
\\
\text{s.t.}\quad &
e_{t,0} = e_{t}^{\mathrm{init}},
\label{eq:reduced-standalone-initial}
\\
&
e_{t,h+1}
=
e_{t,h}
+\eta_{\mathrm{ch}}\Delta u^{\mathrm{ch}}_{t,h}
-\frac{\Delta}{\eta_{\mathrm{dis}}}u^{\mathrm{dis}}_{t,h},
\label{eq:reduced-standalone-dynamics}
\\
&
m^{\mathrm{imp}}_{t,h}-u^{\mathrm{ch}}_{t,h}+u^{\mathrm{dis}}_{t,h}-x^S_{t,h}-x^B_{t,h}-c_{t,h}
=
\hat L_{t,h}-\hat S_{t,h},
\label{eq:reduced-standalone-balance}
\\
&
0\le z_{t,h}\le u^{\mathrm{ch}}_{t,h},
\label{eq:reduced-standalone-z-bounds}
\\
&
0\le x^B_{t,h}\le u^{\mathrm{dis}}_{t,h},
\label{eq:reduced-standalone-xb-bounds}
\\
&
z_{t,h}+x^S_{t,h}+c_{t,h}\le \hat S_{t,h},
\label{eq:reduced-standalone-solar-bounds}
\\
&
m^{\mathrm{imp}}_{t,h}-u^{\mathrm{ch}}_{t,h}+z_{t,h}\ge 0,
\label{eq:reduced-standalone-grid-load-nonneg}
\\
&
0\le e_{t,h}\le E^{\max},
\label{eq:reduced-standalone-energy-bounds}
\\
&
0\le u^{\mathrm{ch}}_{t,h}\le P^{\max}_{\mathrm{ch}},
\label{eq:reduced-standalone-charge-bounds}
\\
&
0\le u^{\mathrm{dis}}_{t,h}\le P^{\max}_{\mathrm{dis}},
\label{eq:reduced-standalone-discharge-bounds}
\\
&
e_{t,h+1}\ge r_{t,h+1},
\label{eq:reduced-standalone-reserve-hard}
\\
&
m^{\mathrm{imp}}_{t,h},u^{\mathrm{ch}}_{t,h},u^{\mathrm{dis}}_{t,h},z_{t,h},x^S_{t,h},x^B_{t,h},c_{t,h}\ge 0.
\label{eq:reduced-standalone-nonneg}
\end{align}
Unless otherwise indicated, the constraints in \eqref{eq:reduced-standalone-dynamics}--\eqref{eq:reduced-standalone-nonneg} are imposed for $h=0,\ldots,H-1$; the energy bounds \eqref{eq:reduced-standalone-energy-bounds} use $h=0,\ldots,H$.

The objective maximizes forecast firm margin over the horizon plus a terminal salvage value for stored energy. Constraints~\eqref{eq:reduced-standalone-initial}--\eqref{eq:reduced-standalone-dynamics} initialize and evolve battery energy. Constraint~\eqref{eq:reduced-standalone-balance} enforces reduced power balance. Constraints~\eqref{eq:reduced-standalone-z-bounds}--\eqref{eq:reduced-standalone-solar-bounds} govern the routing of solar and battery discharge, including metered export through $x^S_{t,h}+x^B_{t,h}$. Constraint~\eqref{eq:reduced-standalone-grid-load-nonneg} ensures that the reduced controls admit a physically meaningful routing of behind-the-meter flows by requiring total grid import to be at least as large as the portion of battery charging supplied by the grid. Constraints~\eqref{eq:reduced-standalone-energy-bounds}--\eqref{eq:reduced-standalone-discharge-bounds} impose battery energy and power limits, and constraint~\eqref{eq:reduced-standalone-reserve-hard} enforces the reserve requirement.

The one-step objective term is
\begin{equation}
\Pi^{\mathrm{SA}}_{t,h}
=
\Delta\Big[
p^{\mathrm{ret}}\hat L_{t,h}
-
\bigl(\hat\lambda_{t,h}+C_{\mathrm{TDSP}}\bigr)m^{\mathrm{imp}}_{t,h}
+
\hat\lambda_{t,h}\bigl(x^S_{t,h}+x^B_{t,h}\bigr)
-
\beta\bigl(z_{t,h}+x^S_{t,h}\bigr)
\Big].
\label{eq:reduced-standalone-objective}
\end{equation}

The point of the variables $z_{t,h}$, $x^S_{t,h}$, and $x^B_{t,h}$ is purely economic: solar used to charge the battery, solar exported to the grid, and battery export face different accounting terms under the tariff. The optimizer therefore keeps a small amount of routing detail needed for the tariff accounting, without introducing a detailed electrical network model.

\subsection{Pooled Dispatch LP}\label{subsec:pooled-mpc}

The pooled controller extends the standalone dispatch LP by allowing internal energy sharing within the fixed pool $\mathcal{G}$. At decision epoch $t$ it keeps the same home-level controls as in the standalone LP,
\[
m^{\mathrm{imp}}_{g,t,h},\quad
u^{\mathrm{ch}}_{g,t,h},\quad
u^{\mathrm{dis}}_{g,t,h},\quad
z_{g,t,h},\quad
x^S_{g,t,h},\quad
x^B_{g,t,h},\quad
c_{g,t,h},
\qquad g\in\mathcal{G},\; h=0,\ldots,H-1,
\]
together with battery energy levels
\[
e_{g,t,h},
\qquad g\in\mathcal{G},\; h=0,\ldots,H,
\]
and adds internal sharing variables
\[
y^S_{g,t,h},\quad
y^B_{g,t,h},\quad
w^L_{g,t,h},\quad
w^C_{g,t,h},
\qquad g\in\mathcal{G},\; h=0,\ldots,H-1.
\]
Here $y^S_{g,t,h}$ is solar sent from home $g$ into the pool, $y^B_{g,t,h}$ is battery discharge sent from home $g$ into the pool, $w^L_{g,t,h}$ is pooled energy received by home $g$ and used to serve load, and $w^C_{g,t,h}$ is pooled energy received by home $g$ and used to charge the battery. When the sharing variables are set to zero, the pooled LP reduces exactly to the standalone LP applied separately to each home. Reserve targets $(r_{g,t,h+1})_{g,h}$ are again treated as exogenous inputs here and are defined in Section~\ref{sec:backup-accounting}.

Given forecasts $(\hat L_{g,t,h},\hat S_{g,t,h},\hat\lambda_{t,h})_{g,h}$ and reserve targets $(r_{g,t,h+1})_{g,h}$, the pooled LP is
\begin{align}
\max\quad &
\sum_{g\in\mathcal{G}}\sum_{h=0}^{H-1}\Pi^{\mathrm{pool}}_{g,t,h}
+
\lambda_{\mathrm{salv},t}\sum_{g\in\mathcal{G}} e_{g,t,H}
\label{eq:pool-master-obj}
\\
\text{s.t.}\quad &
e_{g,t,0}=e^{\mathrm{init}}_{g,t},
\label{eq:pool-initial}
\\
&
e_{g,t,h+1}
=
e_{g,t,h}
+\eta_{g,\mathrm{ch}}\Delta u^{\mathrm{ch}}_{g,t,h}
-\frac{\Delta}{\eta_{g,\mathrm{dis}}}u^{\mathrm{dis}}_{g,t,h},
\label{eq:pool-dynamics}
\\
&
m^{\mathrm{imp}}_{g,t,h}
+w^L_{g,t,h}+w^C_{g,t,h}
-u^{\mathrm{ch}}_{g,t,h}
+u^{\mathrm{dis}}_{g,t,h}
-x^S_{g,t,h}-x^B_{g,t,h}
-y^S_{g,t,h}-y^B_{g,t,h}
-c_{g,t,h}
\notag
\\
&\qquad
=
\hat L_{g,t,h}-\hat S_{g,t,h},
\label{eq:pool-balance}
\\
&
z_{g,t,h}+w^C_{g,t,h}\le u^{\mathrm{ch}}_{g,t,h},
\label{eq:pool-charge-source}
\\
&
0\le x^B_{g,t,h}+y^B_{g,t,h}\le u^{\mathrm{dis}}_{g,t,h},
\label{eq:pool-batt-out-bounds}
\\
&
z_{g,t,h}+x^S_{g,t,h}+y^S_{g,t,h}+c_{g,t,h}\le \hat S_{g,t,h},
\label{eq:pool-solar-bounds}
\\
&
m^{\mathrm{imp}}_{g,t,h}-u^{\mathrm{ch}}_{g,t,h}+z_{g,t,h}+w^C_{g,t,h}\ge 0,
\label{eq:pool-load-nonneg}
\\
&
\sum_{g\in\mathcal{G}}\bigl(w^L_{g,t,h}+w^C_{g,t,h}\bigr)
=
\sum_{g\in\mathcal{G}}\bigl(y^S_{g,t,h}+y^B_{g,t,h}\bigr),
\label{eq:pool-sharing-balance}
\\
&
w^L_{g,t,h}+w^C_{g,t,h}+y^S_{g,t,h}+y^B_{g,t,h}
\le
\sum_{i\in\mathcal{G}}\bigl(y^S_{i,t,h}+y^B_{i,t,h}\bigr),
\label{eq:pool-no-self}
\\
&
0\le e_{g,t,h}\le E_g^{\max},
\label{eq:pool-energy}
\\
&
0\le u^{\mathrm{ch}}_{g,t,h}\le P^{\max}_{g,\mathrm{ch}},
\label{eq:pool-charge}
\\
&
0\le u^{\mathrm{dis}}_{g,t,h}\le P^{\max}_{g,\mathrm{dis}},
\label{eq:pool-discharge}
\\
&
e_{g,t,h+1}\ge r_{g,t,h+1},
\label{eq:pool-reserve-hard}
\\
&
m^{\mathrm{imp}}_{g,t,h},u^{\mathrm{ch}}_{g,t,h},u^{\mathrm{dis}}_{g,t,h},z_{g,t,h},
 x^S_{g,t,h},x^B_{g,t,h},c_{g,t,h},
\notag
\\
&\qquad
y^S_{g,t,h},y^B_{g,t,h},w^L_{g,t,h},w^C_{g,t,h}\ge 0.
\label{eq:pool-nonneg}
\end{align}
Unless otherwise indicated, the home-level constraints in \eqref{eq:pool-dynamics}--\eqref{eq:pool-nonneg} are imposed for all $g\in\mathcal{G}$ and $h=0,\ldots,H-1$; the initial condition \eqref{eq:pool-initial} is imposed for all $g\in\mathcal{G}$, the energy bounds \eqref{eq:pool-energy} use $h=0,\ldots,H$, and the pool-conservation and no-self-sharing constraints \eqref{eq:pool-sharing-balance}--\eqref{eq:pool-no-self} are imposed for $h=0,\ldots,H-1$.

The objective maximizes forecast firm margin over the horizon plus a terminal salvage value for stored energy. Constraints~\eqref{eq:pool-initial}--\eqref{eq:pool-dynamics} initialize and evolve each home's battery energy. Constraint~\eqref{eq:pool-balance} enforces the reduced home-level power balance with internal sharing. Constraint~\eqref{eq:pool-charge-source} decomposes battery charging into local-solar, pooled, and grid-supplied components, while constraints~\eqref{eq:pool-batt-out-bounds}--\eqref{eq:pool-solar-bounds} govern the routing of battery discharge and solar. Constraint~\eqref{eq:pool-load-nonneg} ensures that total grid import is at least as large as the portion of charging supplied by the grid. Constraint~\eqref{eq:pool-sharing-balance} is the pool conservation law, and constraint~\eqref{eq:pool-no-self} prevents a home from receiving its own pooled outflow. Constraints~\eqref{eq:pool-energy}--\eqref{eq:pool-discharge} impose home-specific battery energy and power limits, and constraint~\eqref{eq:pool-reserve-hard} enforces the reserve requirement for each home.

The one-step pooled objective for home $g$ is
\begin{equation}
\Pi^{\mathrm{pool}}_{g,t,h}
=
\Delta\Big[
p^{\mathrm{ret}}\hat L_{g,t,h}
-
\bigl(\hat\lambda_{t,h}+C_{\mathrm{TDSP}}\bigr)m^{\mathrm{imp}}_{g,t,h}
+
\hat\lambda_{t,h}\bigl(x^S_{g,t,h}+x^B_{g,t,h}\bigr)
-
\beta\bigl(z_{g,t,h}+x^S_{g,t,h}+y^S_{g,t,h}\bigr)
\Big].
\label{eq:pool-objective}
\end{equation}

The pooled LP differs from the standalone LP only through the internal sharing variables $y^S_{g,t,h}$, $y^B_{g,t,h}$, $w^L_{g,t,h}$, and $w^C_{g,t,h}$. The variables $y^S_{g,t,h}$ and $y^B_{g,t,h}$ track solar and battery discharge sent into the pool, while $w^L_{g,t,h}$ and $w^C_{g,t,h}$ track pooled energy used at the receiving home for contemporaneous load service and battery charging. Solar sent into the pool, like solar used to charge the home's own battery or exported to the grid, is routed away from direct contemporaneous self-consumption and therefore enters the solar-credit term. When internal sharing is disabled, the model reduces exactly to the standalone dispatch LP applied separately to each home.

\section{Reserve Requirements}\label{sec:backup-accounting}

\paragraph{Forecasting Inputs.}

In the model section, the reserve targets $r_{t,h+1}$ and $r_{g,t,h+1}$ are treated as exogenous MPC inputs. The MPC uses point forecasts for load, solar, and price. The additional input specific to backup service is the reserve requirement, which is constructed from the empirical distribution of forward $T$-hour positive-net-load energy. In the pooled setting, the same procedure is applied home by home to produce home-specific reserve targets; to keep notation light, the formulas below suppress the home index except where needed.

Point forecasts and reserve quantiles are estimated from the raw one-minute Base data using local-time neighborhoods pooled across days. For a target quarter-hour of day $q$, point forecasts pool one-minute observations from all days whose local clock time falls within $\pm k_f$ minutes of $q$, with $k_f=15$. These point forecasts provide the MPC inputs $\hat L$, $\hat S$, and $\hat\lambda$. Reserve quantiles are constructed analogously using a wider local-time neighborhood, with $k_b=30$. For each one-minute observation in that reserve neighborhood, we compute the forward positive-net-load energy over the next $T$ hours; these forward sums form an empirical distribution of $T$-hour outage-energy requirements for that local time, and observations from different days but similar local clock times are treated as exchangeable in constructing that distribution. Price forecasts are built from ERCOT real-time price history using quarter-hour slot medians.

Fix a backup duration $T\in\{2,4,6,8,12,24\}$ hours and let $K(T):=T/\Delta$ be the corresponding number of 15-minute control intervals. Suppressing the home index, define the forward positive-net-load energy requirement over the next $T$ hours starting at reserve time $t+h+1$ by
\begin{equation}
Q^{(T)}_{t,h+1}
:=
\Delta\sum_{j=0}^{K(T)-1}\bigl(N_{t+h+1+j}\bigr)_{+},
\label{eq:forward-positive-netload-energy}
\end{equation}
where $(x)_{+}:=\max\{x,0\}$. Let $q^{(0.90)}_{t,h+1}$ denote the estimated 0.90 quantile of the random variable $Q^{(T)}_{t,h+1}$ under this local-neighborhood construction. Thus $q^{(0.90)}_{t,h+1}$ is the modeled amount of positive net load that must be served over the next $T$ hours, starting at reserve time $t+h+1$, at the 0.90 quantile. In pooling, the same construction is applied home by home, yielding $q^{(0.90)}_{g,t,h+1}$ and hence reserve targets $r_{g,t,h+1}$.
For example, for a 4-hour backup product at a 6pm reserve time, we pool minute-level observations from all days in the sample whose local clock time falls within the 6pm reserve neighborhood, sum each observation's positive net load over the following four hours, take the 90th percentile of those sums, and divide by discharge efficiency to obtain the required stored energy.

\paragraph{Benchmark Reserve Target.}

The benchmark reserve target converts this delivered-energy requirement into required internal battery energy. Because $q^{(0.90)}_{t,h+1}$ measures energy that must be delivered over the outage window while the MPC state $e_{t,h}$ measures internal stored energy, the corresponding reserve requirement is
\begin{equation}
r_{t,h+1}
=
\frac{1}{\eta_{\mathrm{dis}}}
q^{(0.90)}_{t,h+1}
\label{eq:r-main}
\end{equation}
in standalone and, home by home,
\[
r_{g,t,h+1}
=
\frac{1}{\eta_{g,\mathrm{dis}}}
q^{(0.90)}_{g,t,h+1}
\]
in pooling.

This reserve construction is used throughout the paper: it determines each home's longest feasible backup tier in the standalone screening step and defines the reserve floor in the retained-cohort pooling comparisons. Operationally, the backup requirement is implemented as a rolling reserve floor: after every 15-minute control interval, the battery must hold enough internal energy for the promised outage window beginning at that time.

\section{Results}\label{sec:results}

We report weekly firm margin throughout. Its dispatch-dependent component is computed ex post from the implemented trajectory under realized prices using the same tariff conventions as in the control model. We then add pro-rated subscription revenue: \$19 per month for one-battery homes and \$29 per month for two-battery homes, scaled by one quarter for the 7-day experiment. Because this subscription term does not vary with dispatch in the comparisons below, it is omitted from the MPC objective and included only in reported total firm margin.

Thus reported firm margin equals realized dispatch margin summed over the week plus pro-rated subscription revenue; in pooled runs, realized dispatch margin is first summed across homes.

\subsection{Standalone Benchmark}
\label{sec:standalone-results}

We begin by reconstructing the product-eligible sample from the standalone reruns. For each week-complete home and each menu tier $T\in\{2,4,6,8,12,24\}$, a home-tier pair is classified as feasible if the standalone MPC remains feasible at every decision epoch and the realized post-decision state of charge satisfies the tier's reserve floor throughout the observed week. Of the 581 week-complete homes, 38 are infeasible even at the 2-hour tier and are therefore dropped. The remaining 543 homes form the retained cohort. For each retained home $g$, let $T_{g,\mathrm{maxfeas}}$ denote the longest menu tier that home can support in standalone operation. This assignment defines the product each home carries into the benchmark comparison.

The retained cohort is concentrated at shorter products: 149 homes are assigned 2 hours and 174 are assigned 4 hours, whereas only 27 homes qualify for 24 hours. Evaluating each home's matched standalone rerun at its assigned tier yields total weekly standalone firm margin of \$5{,}857.59, or \$10.79 per retained home on average.

\subsection{Pooling on the Retained Cohort}\label{sec:pooling-results}

The pooled benchmark uses exactly the same 543 retained homes and the same home-specific product assignments as the standalone benchmark. The comparison therefore isolates the value of coordination while holding household-level backup promises fixed. In the pooled rerun, the controller can share energy across homes---either to serve contemporaneous load or to charge another home's battery---but it still tracks a separate battery state and a separate reserve requirement for each home.

The longest-feasible assignment also provides the baseline for a simple backup-cap experiment. We impose common caps of 2, 4, 6, 8, 12, and 24 hours. For a given cap $T$, home $g$ is assigned
\[
T_g=\min\{T_{g,\mathrm{maxfeas}},T\}.
\]
Because the menu itself ends at 24 hours, the 24-hour case reproduces the baseline longest-feasible assignment. We then rerun both standalone control and pooling on the same 543 retained homes under these capped assignments. Table~\ref{tab:cap-spectrum} reports the resulting spectrum together with the number of retained homes assigned each cap.

The pattern is clear and monotone. Standalone firm margin is quite stable, declining only from \$11.06 per home at the 2-hour cap to \$10.79 at the 24-hour cap. By contrast, pooling benefit falls steadily as the backup cap rises, from \$1.49 per home at the 2-hour cap to \$1.27 at the 24-hour cap. Relative to standalone firm margin, the value of pooling also declines slightly, from 13.46\% to 11.80\%. Across the full spectrum, pooling contributes a little over one dollar per home-week---about one-eighth of the firm margin earned by serving the home individually.

\begin{table}[t]
\centering
\caption{Pooling across a spectrum of backup caps on the retained cohort. For each cap $T$, retained home $g$ is assigned $T_g=\min\{T_{g,\mathrm{maxfeas}},T\}$, where $T_{g,\mathrm{maxfeas}}$ is the longest backup tier that home can support in standalone operation. The second column counts retained homes at the cap, i.e., homes with $T_g=T$; under the assignment rule above, this is equivalently the number with $T_{g,\mathrm{maxfeas}}\ge T$. The standalone and pooled controllers are both rerun under that same capped assignment, and the 24-hour cap coincides with the baseline longest-feasible benchmark.}
\label{tab:cap-spectrum}
\begin{tabular}{rrrrr}
\toprule
\shortstack{Backup\\cap (h)} & \shortstack{Homes\\at\\cap} & \shortstack{Standalone\\firm margin\\per home (USD)} & \shortstack{Pooling\\benefit\\per home (USD)} & \shortstack{Benefit /\\standalone firm\\margin (\%)}\\
\midrule
2  & 543 & 11.06 & 1.49 & 13.46 \\
4  & 394 & 10.90 & 1.37 & 12.54 \\
6  & 220 & 10.85 & 1.31 & 12.07 \\
8  & 150 & 10.83 & 1.29 & 11.88 \\
12 & 73  & 10.81 & 1.28 & 11.81 \\
24 (= max feasible) & 27 & 10.79 & 1.27 & 11.80 \\
\bottomrule
\end{tabular}
\end{table}

\begin{figure}[t]
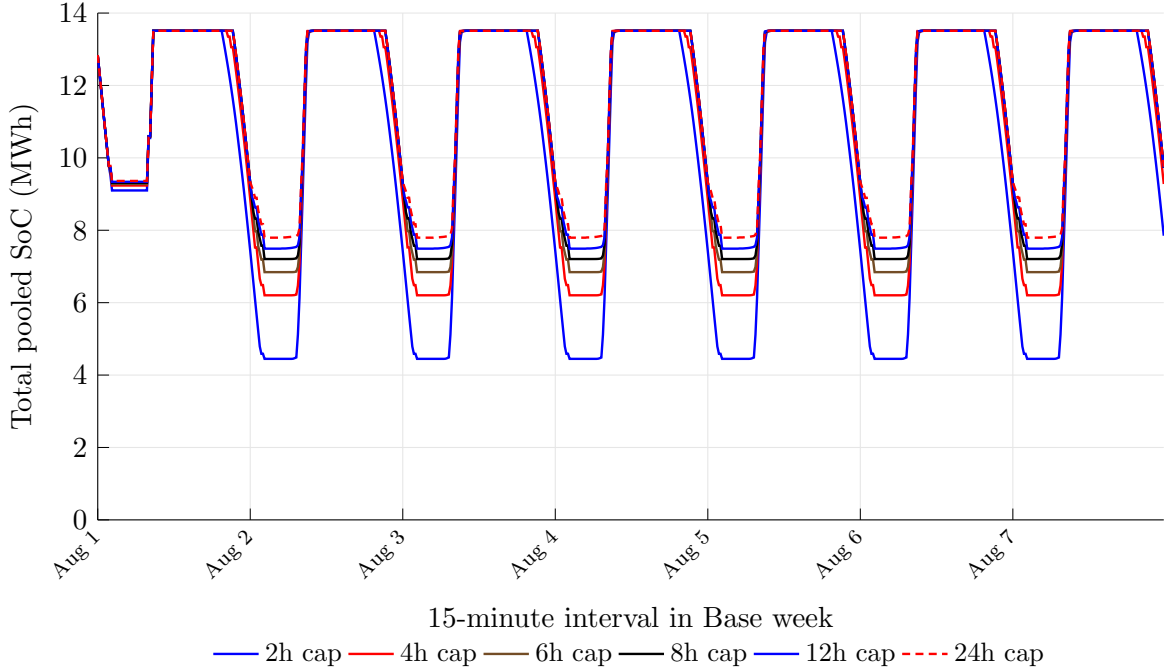

\centering
%
\caption{Total pooled state-of-charge trajectories under the backup-cap experiment. Each curve reports the pooled fleet's total state of charge at the start of each 15-minute interval over the Base week for the capped assignment $T_g=\min\{T_{g,\mathrm{maxfeas}},T\}$. Higher caps keep more energy in the fleet because more battery energy must be reserved to protect household-specific backup commitments.}
\label{fig:pooled-soc-by-cap}
\end{figure}

Figure~\ref{fig:pooled-soc-by-cap} plots the pooled total state of charge over the benchmark week for each backup cap. The pattern is visually consistent with the cap-spectrum comparison in Table~\ref{tab:cap-spectrum}. Higher backup caps keep more energy in the fleet throughout the week, because more battery energy must remain behind household-specific reserve floors. Looser caps allow the pooled controller to draw the fleet down further when doing so is economically attractive.

This monotone decline has a clear operational explanation. As the backup cap rises, each home has to keep more battery energy in reserve in case an outage occurs. That leaves less energy available for flexible economic use during normal operation. In standalone operation, this mainly reduces the home’s ability to shift energy over time to improve its own weekly firm margin, so standalone value declines only slightly across caps. In pooled operation, the loss of flexibility matters more because each home still has to protect its own reserve, so the pool has less energy that can be shifted across homes when their timing, net load, or charging opportunities differ. Because the pooled controller cannot treat all stored energy as one fully shared stock, tighter backup caps reduce the extra value created by coordination and therefore make the pooling gain smaller as the cap increases.

\section{Conclusion}\label{sec:conclusion}

This paper develops an operational framework for residential battery coordination when backup protection is part of the product being sold. Its central conceptual contribution is to identify the right operational object for that problem: once backup commitments are made at the household level, pooling is no longer well represented by a virtual-battery abstraction. Energy can be shared across homes, but battery states and reserve requirements remain household specific, so the controller must keep one battery state and one reserve obligation for each home.

The main empirical message is that pooling remains beneficial under this more realistic structure. Across the backup-cap spectrum we study, pooling adds positive value relative to standalone operation, and that incremental value declines as backup promises become more demanding. More broadly, residential battery fleets should be evaluated as joint economic-and-resilience service systems rather than purely as arbitrage assets. Tighter backup products lock more energy behind home-specific reserve floors and therefore leave less flexibility for coordination gains.

\bibliographystyle{plainnat}
\bibliography{references}

\end{document}